\documentclass[a4paper]{article}

\usepackage[applemac]{inputenc}

\usepackage{a4wide}

\newtheorem{theorem}{Theorem}[section]

\newtheorem{proposition}[theorem]{Proposition}
\newtheorem{lemma}[theorem]{Lemma}
\newtheorem{corollary}[theorem]{Corollary}
\newtheorem{remark}[theorem]{Remark}
\newenvironment{proof}{\pagebreak[3]\noindent\textit{Proof: }}{\pagebreak[3]\medskip}
\newtheorem{expl}[theorem]{Example}
\newenvironment{example}[1][]{\ifthenelse{\equal{#1}{}}{\begin{expl}\upshape}{\begin{expl}[#1]\upshape}}{\hspace*{\fill}$\Diamond$\pagebreak[3]\par\end{expl}}
\newcommand{\qed}{\hspace*{\fill}$\Box$
\pagebreak[3]\medskip}

\usepackage{latexsym}
\usepackage{amssymb}
\usepackage{amsmath}
\usepackage{enumerate,xspace}
\usepackage{ifthen}
\usepackage{booktabs}
\usepackage{tikz}
\usepackage[colorlinks,citecolor=blue,final]{hyperref}
\usepackage{multirow}
\usepackage{subfigure}

\newenvironment{vd}{\noindent\color{blue} VD }{}

\newenvironment{mk}{\noindent\color{green} MK }{}

\newcommand{\set}[2]{\left\{#1\mathrel{\left|\vphantom{#1}\vphantom{#2}\right.}#2\right\}}
\newcommand{\oneset}[1]{\left\{\mathinner{#1}\right\}}
\newcommand{\smallset}[1]{\left\{\mathinner{#1}\right\}}

\newcommand{\abs}[1]{\left|\mathinner{#1}\right|}
\newcommand{\N}{\mathbb{N}}

\newcommand{\gR}{\mathrel{\mathcal{R}}}

\newcommand{\gL}{\mathrel{\mathcal{L}}}

\newcommand{\gJ}{\mathrel{\mathcal{J}}}

\newcommand{\gD}{\mathrel{\mathcal{D}}}

\newcommand{\gRop}{\mathbin{\gR}}

\newcommand{\FO}{\mathrm{FO}}  %%% first order logic
\newcommand{\BFO}{\mathrm{\mathbf{FO}}}

\let\tilde\widetilde

\newcommand{\SU}{\mathbin{\mathsf{S\kern-0.1emU}}}

\newcommand{\DA}{\mbox{\bf DA}}

\newcommand{\Synt}{\mathrm{Synt}}

\colorlet{DKbackcolora}{cyan!5}
\colorlet{DKbackcolorA}{cyan!12}
\colorlet{DKbackcolorB}{cyan!22}
\colorlet{DKbackcolorC}{cyan!32}
\colorlet{DKbackcolorD}{cyan!42}
\colorlet{DKbackcolorE}{cyan!52}
\colorlet{DKbackcolorF}{cyan!62}
\colorlet{DKfontcolor}{black}
\newcommand{\mydot}{\color{DKfontcolor}\bullet}
\newcommand{\A}{\mathcal{A}}

\newcommand{\ov}[1]{ \overline{#1}}
\newcommand{\alp}{\mathop{\mathrm{alph}}}
\newcommand{\im}{\mathop{\mathrm{im}}}
\newcommand{\EF}{Ehren\-feucht-Fra{\"{\i}}ss{\'e}}

\newcommand{\IFF}{if and only if\xspace}
\newcommand{\codet}{complement-deter\-mi\-nistic\xspace}
\newcommand{\arr}{arrow\xspace}
\newcommand{\e}{1}

\renewcommand{\phi}{\varphi}

\renewcommand{\implies}{\ \Rightarrow \ }

\newcommand{\Sg}{\Sigma}
\newcommand{\GA}{\Gamma}
\newcommand{\oo}{\omega}
\newcommand{\cpx}[1]{{#1}^\infty \cap {#1}^{\im}}
\begin{document}

%\iffalse %draft
\title{Fragments of first-order logic over infinite words}

\author{Volker Diekert \qquad Manfred Kuf\-leitner \\[5mm]
  Universit{\"a}t Stuttgart, FMI \\
  Universit{\"a}tsstra{\ss}e 38 \\
  D-70569 Stuttgart, Germany \\[5mm]
  \texttt{$\{$diekert$,$kufleitner$\}$@fmi.uni-stuttgart.de}}

\date{October 2nd, 2009} % optional

%\keywords{infinite words, regular languages, first-order logic,
%  automata theory, semigroups, topology}

%\subjclass{F.4.1 Mathematical Logic, F.4.3 Formal Languages.}

\maketitle
%\fi %draft

{%\footnotesize
\tableofcontents
}

\pagebreak

\begin{abstract}
  \noindent
  \textbf{Abstract.} \ 
  We give topological and algebraic characterizations as well as
  language theoretic descriptions of the following subclasses of
  first-order logic $\FO[{<}]$ for $\omega$-languages: $\Sigma_2$,
  $\FO^2$, $\FO^2 \cap \Sigma_2$, and $\Delta_2$ (and by duality
  $\Pi_2$ and $\FO^2 \cap \Pi_2$). These descriptions extend the
  respective results for finite words. In particular, we relate the
  above fragments to language classes of certain (unambiguous)
  polynomials. An immediate consequence is the decidability of the
  membership problem of these classes, but this was shown before by
  Wilke~\cite{wil98} and Boja{\'n}czyk~\cite{boj08fossacs} and is
  therefore not our main focus. The paper is about the interplay of
  algebraic, topological, and language theoretic properties.
\end{abstract}

\section{Introduction}

The algebraic approach is fundamental for the understanding of regular
languages. It has been particularly fruitful for fragments of
first-order logic over finite words.  For example, a result of Wilke
and Th{\'e}rien is that $\FO^2$ and $\Delta_2$ have the same
expressive power~\cite{tw98stoc}, where the latter class by definition
denotes $\Sg_2 \cap \Pi_2$. Further results are language theoretic and
(very often decidable) algebraic characterizations of logical
fragments, see e.g.~\cite{tt07lmcs} or~\cite{dgk08ijfcs} for
surveys. Several results for finite words have been extended to other
structures such as trees and other graphs, see~\cite{wei04} for a
survey.
More recently, $\FO^2$, $\Delta_2$, and $\Sigma_2$ have been
characterized for Mazurkiewicz traces~\cite{dhk07mazu,kuf06dlt}; 
$\Delta_2$ and the Boolean closure of $\Sigma_1$ have been
characterized for unranked trees~\cite{bs08icalp,bss08lics}.
For some characterizations over finite words, it has been shown that
they cannot be generalized; e.g.\ over unranked trees, it turned out
that $\FO^2$ and $\Delta_2$ are incomparable~\cite{boj07lics}.  For
infinite words, the expressive power of $\FO^2$ is not equal to
$\Delta_2$, since saying that letters $a$ and $b$ appear infinitely
often, but $c$ only finitely many times is $\FO^2$-definable, but
there is neither a $\Sg_2$-formula nor a $\Pi_2$-formula specifying
this language.

\begin{figure}[ht]
  \centering
  \subfigure[Finite words]{
  \begin{tikzpicture}[scale=0.55]
    \draw[thin] (0,0) -- (12,0);
    \draw[thin,fill=DKbackcolorA] (0,0) arc (180:0:4cm);
    \draw[thin,fill=DKbackcolorA] (12,0) arc (0:180:4cm);

    \begin{scope}
      \clip (0,0) arc (180:0:4cm);
      \draw[thin,fill=DKbackcolorC] (12,0) arc (0:180:4cm);
    \end{scope}
    
    \draw[thick] (0,0) -- (12,0);
    \draw[thick] (0,0) arc (180:0:4cm);
    \draw[thick] (12,0) arc (0:180:4cm);

    \draw (1.25,0.35) node[above] {$\mathbf{\Sigma_2}$};
    \draw (10.75,0.35) node[above] {$\mathbf{\Pi_2}$};
    \draw (6,0.35) node[above] {$\mathbf{\Delta_2 = \BFO^2}$};
  \end{tikzpicture}
  \label{figfin}
  } \quad
  \subfigure[Finite and infinite words]{
  \begin{tikzpicture}[scale=0.55]
    \draw[thin] (0,0) -- (12,0);
    \draw[thin,fill=DKbackcolorA] (0,0) arc (180:0:4cm);
    \draw[thin,fill=DKbackcolorA] (12,0) arc (0:180:4cm);
    \draw[thin,fill=DKbackcolorA] (2.5,0) arc (180:0:3.5cm and 6cm);
    
    \begin{scope}
      \clip (2.5,0) arc (180:0:3.5cm and 6cm);
      \draw[thin,fill=DKbackcolorB] (0,0) arc (180:0:4cm);
      \draw[thin,fill=DKbackcolorB] (12,0) arc (0:180:4cm);
    \end{scope}
    
    \begin{scope}
      \clip (0,0) arc (180:0:4cm);
      \draw[thin,fill=DKbackcolorC] (12,0) arc (0:180:4cm);
    \end{scope}
    
    \draw[thick] (0,0) -- (12,0);
    \draw[thick] (0,0) arc (180:0:4cm);
    \draw[thick] (12,0) arc (0:180:4cm);
    \draw[thick] (2.5,0) arc (180:0:3.5cm and 6cm);

    \draw (1.25,0.35) node[above] {$\mathbf{\Sigma_2}$};
    \draw (10.75,0.35) node[above] {$\mathbf{\Pi_2}$};
    \draw (6,0.35) node[above] {$\mathbf{\Delta_2}$};
    \draw (6,5.8) node[below] {$\mathbf{\BFO^2}$};

    \draw (3.75,2.05) node (L2) {$\mydot$};
    \draw (L2) node[above] {$\color{DKfontcolor}\Gamma^*$};

    \draw (6,3.95) node (L3) {$\mydot$};
    \draw (L3) node[above] {$\color{DKfontcolor}A^{\im}$};

    \draw (8.25,2.05) node (L4) {$\mydot$};
    \draw (L4) node[above] {$\color{DKfontcolor}\Gamma^\oo$};

    \draw (6,2.05) node (L6) {$\mydot$};
    \draw (L6) node[above] {$\color{DKfontcolor}\Gamma^\infty$};
  \end{tikzpicture}
  \label{figotto}
  }
  \caption{The fragments $\Sigma_2$, $\Pi_2$, and $\FO^2$ over finite
    and over finite and infinite words}
\end{figure}
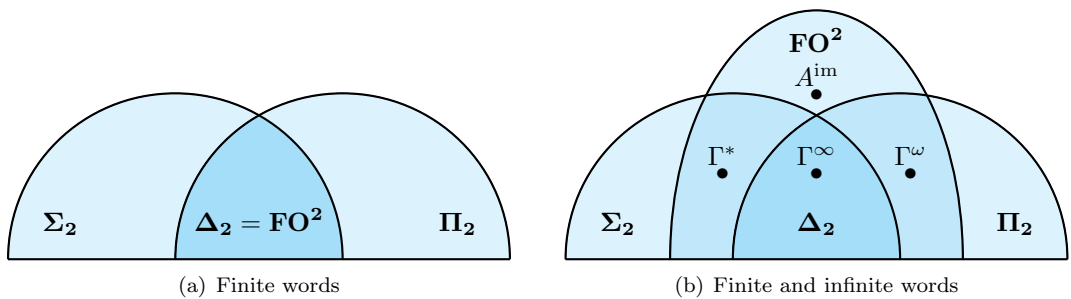

The results about finite words do not translate directly to infinite
words as neither $\Sigma_2$ nor $\Pi_2$ copes with the exact
alphabetic information which letters appear infinitely often,
see Figures~\ref{figfin} and~\ref{figotto}.

Our results deepen the understanding of first-order fragments over
infinite words. A decidable characterization of the membership problem
for $\FO^2$ over infinite words has been given in the habilitation
thesis of Wilke~\cite{wil98}. Recently, decidability for $\Sigma_2$
has been shown independently by Boja{\'n}czyk
\cite{boj08fossacs}. Language theoretic and decidable algebraic
characterizations of the fragment $\Sigma_1$ and of its Boolean closure
can be found in~\cite{pp04,Pin98a}.

We introduce two generalizations of the usual Cantor topology for
infinite words.  One of our first results is a characterization of
$\Sigma_2$-definability for languages in $\Gamma^\infty$.  This
characterization consists of two components: The first one is an
algebraic property of the syntactic monoid and the second part is
requiring that $L$ is open in some alphabetic topology. Both
properties are decidable.

Our second result is that a regular language is $\FO^2$-definable if
and only if its syntactic monoid is in the variety $\DA$. (The result
is surprising in the sense that it contradicts a statement in
\cite{wil98}). In addition, we show that a language is definable in
$\FO^2$ if and only if it is closed in some further refined alphabetic
topology and if it is weakly recognizable by a monoid in $\DA$.  In
particular, weak recognition and strong recognition do not coincide
for the variety $\DA$. This seems to be a new result as well. We also
contribute a language theoretic characterization of $\FO^2$ in terms
of unambiguous polynomials with additional constraints on the letters
which occur infinitely often.

Other results of our paper are the characterization of $\FO^2 \cap
\Sigma_2$ as the class of unambiguous polynomials and of $\Delta_2$ in
terms of unambiguous polynomials in some special form and also in
terms of deterministic languages. It follows already from this
description that $\Delta_2$ is a proper subset of
$\FO^2$. Furthermore, we show that the equality of $\FO^2$ and
$\Delta_2$ holds relativized to some fixed set of letters which occur
infinitely often. If this set of letters is empty, we obtain the
situation for finite words as a special case. Finally, we relate
topological constructions such as \emph{interior} and \emph{closure}
with membership in the fragments under consideration.  A brief summary
of the results for the various fragments can be found in
Section~\ref{sec:summary} at the end of this paper.

For basic notions on languages of infinite words we refer to standard
references such as~\cite{pp04,tho90handbook}.  Most results of the
present paper are from its conference version~\cite{dk09stacs}, but
for lack of space they appeared in many cases without proof.  The
present journal version gives full proofs and some new material. In
particular, we give a new characterization of $\oo$-regular
$\Delta_2$-languages involving deterministic and \codet languages,
cf.\ Corollary~\ref{cor:d2o}.

%%%%%%%%%%%%%%%%%%%%%%%%%%%%%%%%%%%%%%%%%%

\section{Preliminaries}\label{sec:pre}

\paragraph{Words}
Throughout, $\Gamma$ is a finite alphabet, $A \subseteq \Gamma$ is a
subset of the alphabet, $u,v,w$ are finite words, and
$\alpha,\beta,\gamma$ are finite or infinite words. If not specified
otherwise, then in all examples we assume that $\GA$ has three
different letters $a,b,c$.  By $u \leq \alpha$ we mean that $u$ is a
prefix of $\alpha$. By $\alp(\alpha)$ we denote the \emph{alphabet} of
$\alpha$, i.e., the letters occurring in the sequence $\alpha$.  As
usual, $\Gamma^*$ is the free monoid of finite words over $\Gamma$.
The neutral element is the empty word $\e$. If $L$ is a subset of a
monoid, then $L^*$ is the submonoid generated by $L$. For $L \subseteq
\Gamma^*$ we let $L^{\omega} = \set{u_1u_2\cdots}{u_i \in L \text{ for
    all } i \geq 1}$ be the set of infinite products.  We also let
$L^{\infty} = L^* \cup L^{\omega}$.  A natural convention is $\e^\oo =
\e$. Thus, $L^{\infty} = L^{\omega}$ if and only if $\e \in L$.

We write $\im(\alpha)$ for those letters in $\alp(\alpha)$ which have
infinitely many occurrences in $\alpha$. The notation has been
introduced in the framework of so called \emph{complex traces}, see
e.g.~\cite{gp95} for a detailed discussion of this concept.  The
notation $\im(\alpha)$ refers to the \emph{imaginary part} and we
adopt it here, but for our purpose it might be also convenient to
remember $\im(\alpha)$ as an abbreviation for letters which appear
\emph{infinitely many} times in $\alpha$.  Sets of the form $A^{\im}$
play a crucial role in our paper. By definition, $A^{\im}$ is the set
of words $\alpha$ such that $\im(\alpha) = A$. Note that $\Gamma^* =
\emptyset^{\im}$. The set $\Gamma^\infty$ is the disjoint union over
all $A^{\im}$.

\paragraph{Logic and regular sets}
We assume that the reader is familiar with basic concepts in formal
language theory.  Our focus is on \emph{regular} languages. If $L
\subseteq \GA^\infty$ is regular, then we may think that its finitary
part $L \cap \GA^*$ is specified by some NFA and that its infinitary
part $L \cap \GA^\oo$ is specified by some B\"uchi automaton.  For a
unified model to accept regular languages in $\GA^\infty$ it is
convenient to consider an \emph{extended B\"uchi automaton} which has
a finite set of states $Q$ and two types of accepting states, a set of
\emph{final} states $F\subseteq Q$ for accepting finite words and a
set of \emph{repeated} states $R\subseteq Q$ for accepting infinite
words. Thus, this model yields also a natural definition of
\emph{deterministic regular} languages in $\Gamma^\infty$, see below
for more details.

We focus on regular languages which are given by first-order sentences
in $\FO[<]$. Thus, atomic predicates are $\lambda(x)=a$ and $x<y$
saying that position $x$ in a word $\alpha$ is labeled with $a \in
\GA$ and position $x$ is smaller than $y$, respectively.  By $\FO^2$ we
mean $\FO[<]$-sentences which use at most two names $x$ and $y$ as
variables or the class of languages specified by such formulas.
It is well-known that three variables are sufficient to express
any $\FO[<]$-property (see e.g.~\cite{dg08SIWT}), whereas $\FO^2$ is a
proper subclass. 
Similarly, $\Sg_2$ means $\FO[<]$-sentences which are in prenex normal
form and which start with a block of existential quantifiers, followed
by a block of universal quantifiers and a Boolean combination of
atomic formulas. A $\Pi_2$-formula means a negation of a
$\Sg_2$-formula. The notations $\Sg_2$ and $\Pi_2$ refer also to the
corresponding language classes.  The class $\Delta_2$ means the class
of $\Sigma_2$-formulas which have an equivalent $\Pi_2$-formula. But
the notion of equivalence depends on the set of models we use.

If the models are finite words, then a result of Th{\'e}rien and Wilke
\cite{tw98stoc} states $\FO^2 = \Delta_2$. Moreover, $\FO^2$ is the
class of regular languages in $\GA^*$ which are recognized by some
finite monoid in the variety $\DA$ and a classical result of
Sch{\"u}tzenberger shows that $\DA$ also coincides with unambiguous
polynomials~\cite{sch76}.  The variety $\DA$ has been baptized this
way because it means \emph{$\gD$-classes are aperiodic}. More
precisely, $\DA$ contains those finite monoids, where all regular
$\gD$-classes are aperiodic semigroups. We refer to
\cite{tt02,dgk08ijfcs} for more background on the class $\DA$. It is
also the class of finite monoids defined e.g.\ by equations of type
$(xy)^\oo = (xy)^\oo y (xy)^\oo$. Another characterization says that
$\DA$ is defined by finite monoids $M$ satisfying $e=ese$ for all
idempotents $e$ (i.e., $e^2 = e$) and for all $s= s_1\cdots s_n$ where
$e \in Ms_iM$ for each $i$, see
e.g.~\cite{bf84dm,pst88,weil93sf}. This is the definition which we use
below.

Saying that formulas are equivalent if they agree on all finite and
infinite words refines the notion of equivalence for formulas and
changes the picture. This is actually the starting point of this work.
So, in this paper models are finite and infinite words. We are mainly
interested in infinite words, but it does no harm to include finite
words, and this makes the situation more uniform and the results on
finite words reappear as special cases.  See e.g.\
Theorem~\ref{thm:lohengrin} which impies that $\FO^2 = \Delta_2$ for
finite words by choosing $A = \emptyset$.
%%%%%%%%

\paragraph{Recognizability by finite monoids}
By $M$ we denote a finite monoid. We always assume that $M$ is
equipped with a partial order $\leq$ being compatible with the
multiplication, i.e., $u \leq v$ implies $sut \leq svt$ for all
$s,t,u,v \in M$. If not specified otherwise, we may choose $\leq$ to
be the identity relation.

For an idempotent element $e \in M$ we define $M_e = \set{s \in M}{e
  \in MsM}^*$, i.e., $M_e$ is the submonoid of $M$ which is generated
by factors of $e$. If $M$ has a generating set $\Gamma$, then $M_e$ is
generated by $\set{a \in \Gamma}{e \in MaM}$. We can think of this set
as the maximal alphabet of the idempotent $e$.  We say that an
idempotent $e$ is \emph{locally top} (\emph{locally bottom}, resp.) if
$ese \leq e$ ($ese \geq e$, resp.) for all $s \in M_e$.
By $\DA$ we denote the class of finite monoids such that $ese = e$ for
all idempotents $e \in M$ and all $s \in M_e$. Thus, it is the 
class of finite monoids where idempotents are locally top and
locally bottom. 

\begin{remark}\label{rem:sollenwirdiesbemerken}
  Assume that $M$ is generated $\Gamma$. In order to test that $M \in
  \DA$, it is enough to check for all $e = e^2 \in M$ and all $a \in
  \Gamma$ with $e \in MaM$ that we have $eae = e$.  Indeed, consider
  $s \in M_e$ and $a \in \Gamma$ with $e \in MaM$. By induction $ese =
  e$, and it is enough to see that $esae = e$. Now, $ese = e$ implies
  that the element $es$ is idempotent and we have $es \in MaM$, too.
  The result follows:
  \begin{equation*}
    esae = esaese = es es e = e.
  \end{equation*}
\end{remark}

\begin{example}\label{ex:damon}
  Let $M=\oneset{1,a,b,c,ba, 0}$ be the monoid having the following
  description: All elements are idempotent except for $ba$. We have
  $(ba)^2 = ab = 0$, and $0$ behaves like a zero, i.e., $0x=x0=0$ for
  all $x$.  Moreover, we have the equations:
  \begin{equation*}
    ca = a, \  ac = c, \ cb = c, \ bc = b, \ ab = 0.
  \end{equation*}
  The monoid $M$ is not in $\DA$, because $a^2 = a = cba \in MbM$, but
  $aba = 0 \neq a$.  However, the submonoid $N =
  M\setminus\smallset{c}$ is in $\DA$. Visual representations of $M$
  and $N$ in terms of so-called egg-box diagrams (see
  e.g.~\cite{pin86}) can be found in Figures~\ref{fig:exdamonM}
  and~\ref{fig:exdamonN}.
\end{example}

Let $L \subseteq \Gamma^{\infty}$ be a language. The \emph{syntactic
  preorder} ${\leq_L}$ over $\Gamma^*$ is defined as follows. We let
$u \leq_L v$ if for all $x,y,z \in \Gamma^*$ we have both
implications:
\begin{align*}
  xvyz^{\omega} \in L \ \Rightarrow\ xuyz^{\omega} \in L 
  \qquad \text{and} \qquad
  x(vy)^{\omega} \in L \ \Rightarrow\ x(uy)^{\omega} \in L.
\end{align*}
Let us recall that $\e^{\omega} = \e$. Two words $u,v \in \Gamma^*$
are syntactically equivalent, written as $u \equiv_L v$, if both $u
\leq_L v$ and $v \leq_L u$. This is a congruence and the congruence
classes $[u]_L = \set{v \in \Gamma^*}{u \equiv_L v}$ form the
\emph{syntactic monoid} $\Synt(L)$ of $L$. The preorder $\leq_L$ on
words induces a partial order $\leq_L$ on congruence classes, and
$(\Synt(L),{\leq_L})$ becomes an ordered monoid. It is a well-known
classical result that the syntactic monoid of a regular language $L
\subseteq \Gamma^{\infty}$ is finite, see
e.g.~\cite{pp04,tho90handbook}. Moreover, in this case $L$ can be
written as a finite union of languages of type $[u]_L\,
[v]_L^{\omega}$ where $u,v \in \Gamma^*$ with $uv \equiv_L u$ and $v^2
\equiv_L v$. In contrast to finite words, there exist non-regular
languages in $\Gamma^\infty$ with a finite syntactic monoid.

Now, let $h : \Gamma^* \to M$ be any surjective homomorphism onto a
finite ordered monoid $M$ and let $L \subseteq \Gamma^{\infty}$. If
the reference to $h$ is clear, then we denote by $[s]$ the set of
finite words $h^{-1}(s)$ for $s \in M$. We use the following
terminology.
\begin{itemize}
\item $(s,e) \in M \times M$ is a \emph{linked pair}, if $se = s$ and
  $e^2 = e$.
\item $h$ \emph{weakly recognizes} $L$, if 
  \begin{equation*}
    L = \bigcup \set{[s][e]^{\omega}}{(s,e) 
      \text{ is a linked pair and } [s][e]^{\omega} \subseteq L}
  \end{equation*}
\item $h$ \emph{strongly recognizes} $L$ (or simply \emph{recognizes}
  $L$), if
  \begin{equation*}
    L = \bigcup \set{[s][e]^{\omega}}{(s,e) 
      \text{ is a linked pair and } [s][e]^{\omega} \cap L \neq \emptyset}
  \end{equation*}
\item $L$ is \emph{downward closed (on finite prefixes)} for $h$, if
  $[s][e]^{\omega} \subseteq L$ implies $[t][e]^{\omega} \subseteq L$
  for all $s,t,e \in M$ where $t \leq s$.
\end{itemize}

If $L$ is regular, then the syntactic homomorphism $h_L$ strongly
recognizes $L$.

\begin{example}\label{ex:syn}
  Let $\Gamma = \oneset{a,b,c}$ and $L$ be one of the languages
  $\Gamma^*ab\,\Gamma^*$, $\Gamma^*ab\,\Gamma^\oo$, or
  $\Gamma^*ab\,\Gamma^\infty$. The syntactic monoid of $L$ is always
  the same. It has six elements and can be identified with the monoid
  $M=\oneset{1,a,b,c,ba, 0}$ defined in Example~\ref{ex:damon} such
  that the syntactic homomorphism maps $\Gamma$ to the respective
  generators of $M$. Actually we have:
  \begin{align*}
    \Gamma^*ab\,\Gamma^*    &=  [0] = [0][1]^{\omega},\\   
    \Gamma^*ab\,\Gamma^\oo   &= 
    \bigcup \set{[0][e]^{\omega}}{1\neq e \in M}, \\ 
    \Gamma^*ab\,\Gamma^\infty &= \bigcup \set{[0][e]^{\omega}}{ e \in M}.
  \end{align*}
  All of the above languages are strongly recognized by $M$ (using the
  syntactic homomorphism). The language $[0][a]^\omega$ is weakly
  recognized by $M$, but it is not strongly recognized because
  $ab(cbca)^\omega = abc(bcac)^\omega \in [0][a]^\omega \cap
  [0][b]^\omega$ and $ab^\omega \in [0][b]^\omega \setminus
  [0][a]^\omega$.
\end{example}

\begin{lemma}\label{lem:SyntL:DownClosed}
  Let $L \subseteq \Gamma^{\infty}$ be a regular language and let $h_L
  : \Gamma^* \to \Synt(L)$ be its syntactic homomorphism.  Then for
  all $s,t,e,f \in M$ such that $t \leq s$, $f \leq e$, and
  $[s][e]^{\omega} \subseteq L$ we have $[t][f]^{\omega} \subseteq L$.
  In particular, $L$ is downward closed (on finite prefixes) for $h_L$.
\end{lemma}

\proof
Let $u \in [s]$, $x \in [e]$ and let $v \in [t]$, $y \in [f]$. Now,
$ux^{\omega} \in L$ implies $vx^{\omega} \in L$, which in turn implies
$vy^{\omega} \in L$. Since $L$ is regular, $h_L$ strongly recognizes
$L$; and we obtain $[t][f]^{\omega} \subseteq L$, because $vy^{\omega}
\in [t][f]^{\omega}\cap L$.
\qed

\paragraph{Deterministic, \codet, and \arr languages}
Intuitively, the best way to define \emph{deterministic languages} is
to say that a language is \emph{deterministic}, if it is recognized by
a deterministic extended B\"uchi automaton with final and repeated
states as described above.
% vd selbstreferenz, oder? in the preliminaries.  
Therefore, a regular language $L \subseteq
\Gamma^\infty$ is deterministic \IFF its $\oo$-regular part $L \cap
\Gamma^\oo$ can be accepted by some deterministic B\"uchi automaton in
the usual sense. 

There is also a well-known tight connection to what we call here
\emph{\arr languages} $\overrightarrow{W}$: For $W \subseteq \Gamma^*$
we define
\begin{equation*}
  \overrightarrow{W} =  \set{\alpha \in   \Gamma^{\infty}}
  {\text{for every prefix } u \leq \alpha
  \text{ there exists } uv \leq \alpha \text{ with }
   uv \in W}. 
%{\forall u \leq \alpha \, \exists v \colon uv \leq \alpha \wedge uv \in W}.
\end{equation*}

Using B\"uchi automata, we see that a regular language $L \subseteq
\Gamma^\infty$ is deterministic \IFF we can write $L \cap \GA^\oo =
\overrightarrow{W} \cap \GA^\oo$ for some regular $W \subseteq
\Gamma^*$. Actually, a classical result of Landweber yields a more
precise statement: If $L \subseteq \Gamma^\oo$ is $\oo$-regular and $L
= \overrightarrow{W} \cap \GA^\oo$ for some set $W\subseteq \Gamma^*$,
then $W$ can be chosen to be regular, too (which means $L$ is
deterministic) see e.g.~\cite{tho90handbook}. Therefore it is
justified to take the weakest condition as a formal definition
here. Moreover, as we have not formally defined B\"uchi automata, we
use the Landweber characterization as our working definition: If we
speak about a \emph{deterministic} language then we are content with
$L$ being regular and $L \cap \GA^\oo = \overrightarrow{W} \cap
\GA^\oo$ for some set $W \subseteq \Gamma^*$.  It is called
\emph{\codet,} if $\Gamma^\infty\setminus L$ is deterministic. It is
well-known and easy to see (e.g.\ with our working definition) that
deterministic languages are closed under finite union and finite
intersection.

For example, if $W = \GA^*a$, then $\overrightarrow{W} \cap
\Gamma^{\omega}$ is the deterministic $\oo$-regular language of words
having infinitely many $a$'s. Its complement is not deterministic (if
$\abs \GA \geq 2$). Hence \emph{infinitely many $a$'s} is not
\codet. In particular, deterministic languages do not form a Boolean
algebra, whereas the class of languages which are simultaneously
deterministic and \codet does.  Note that the class of \arr languages
is not closed under finite intersection: $\overrightarrow{\GA^*a} \cap
\overrightarrow{\GA^*b}$ is deterministic but not an \arr language (in
our sense) because the intersection is not empty, e.g., it contains
$(ab)^\oo$, but it does not contain any finite word.
 
Our definitions differ slightly from the notation used elsewhere,
where $\overrightarrow{W}$ is commonly used as the $\oo$-language of
those infinite words with infinitely many prefixes in $W$, which is
the set $\overrightarrow{W} \cap \GA^\oo$ in our notation.  In our
definition we have however a closure operator: $W \subseteq
\overrightarrow{W} = W\cup (\overrightarrow{W} \cap \Gamma^{\omega})$.
Moreover, the characterization of $\Delta_2$-languages is more natural
in our definition.  Also note that if $L =\overrightarrow{W}$, then $W
= L \cap \GA^*$. If we only have $L \cap \GA^\oo
=\overrightarrow{W}\cap \GA^\oo$, then there are uncountably many
choices for $W$, in general.

%%%%%%%%%%%%%%%%%%%%%%%%%%%%%%%%%%%%%%%%%%%%%%%%%%%%%
\paragraph{Finite $\mathbf{\omega}$-semigroups} %\label{sec:osg}

The notion of an $\oo$-semigroup has been introduced as a tool for
language varieties of finite and infinite words; and it leads, in
particular, to an Eilenberg-type theorem, see \cite{pp04,wil91icalp}.
Finite $\mathbf{\omega}$-semigroups yield another possible framework
to express most of our results.  Our focus is however to transfer
results from finite words to infinite words using topology, so the
classical theory of recognition by finite monoids turned out to be
suitable for our purposes. But still it might be useful for a possible
generalization to convert our results to the terminology of
$\oo$-semigroups. We refer to the textbook \cite{pp04}, where the
theory has been nicely presented in detail.

\section{The alphabetic topology and polynomials}\label{sec:alphtop}

Topological information is crucial in our characterization results.
Recall that a \emph{topology} on a set $X$ is given by a family of
subsets (called \emph{open subsets}) such that a finite intersection
and an arbitrary union of open subsets is open.  We define the
\emph{alphabetic topology} on the set $\Gamma^\infty$ by its basis,
which is given by all sets of the form $u A^{\infty}$ for $u \in
\Gamma^*$ and $A \subseteq \Gamma$. Thus, a set $L\subseteq
\Gamma^\infty$ is \emph{open} if and only if for each $A\subseteq
\Gamma$ there is a set of finite words $W_{\!A} \subseteq \Gamma^*$
such that $L = \bigcup\, {W_{\!A}}\, A^{\infty}$. By definition, a set
is \emph{closed}, if its complement is open; and it is \emph{clopen},
if it is both open and closed.  For example, the sets $u A^{\infty}$
are clopen.  In particular, the sets $A^{\infty}$ are clopen, too.  A
set of the form $A^{\im}$ is not open unless $A = \emptyset$, it is
not closed unless $A = \Gamma$.

Note that in the alphabetic topology every singleton $u\in \GA^*$ is
open since $u \emptyset^{\infty} = u \oneset{1} =\oneset{u}$.  Thus,
$\Gamma^*$ is an open, discrete, and dense subset of
$\Gamma^\infty$. The alphabetic topology is a refinement of the usual
Cantor topology, where the languages $\oneset{u}$ and $u
\Gamma^{\infty}$ form a basis of (Cantor-)open subsets for $u\in
\GA$. The Cantor space $\Gamma^{\infty}$ is compact.  As soon as
$\Gamma$ has at least two letters more sets are open in the alphabetic
topology than in the Cantor topology.  For example, the sets $u
A^{\infty}$ being clopen in the alphabetic topology are neither open
nor closed in the Cantor topology for $\emptyset \neq A \neq \Gamma$.

\begin{remark}\label{rem:nixcompact}
  The space $\Gamma^\infty$ with the alphabetic topology is Hausdorff.
  It is compact \IFF $\abs{\GA} \leq 1$.  To see that it is not
  compact for $\Gamma= \oneset{a,b}$ note that $\Gamma^\infty $ is
  covered by $a^\infty$ together with open sets of the form
  $ub\Gamma^\infty$ with $u\in \GA^*$. But for no finite subset $F
  \subseteq \Gamma^*$ do we have $\Gamma^\infty = a^\infty \cup
  Fb\Gamma^\infty$.
\end{remark}

For a language $L$, its \emph{closure} $\overline{L}$ is the
intersection of all closed sets containing $L$. A word $\alpha \in
\Gamma^{\infty}$ belongs to $\overline{L}$ if for all open subsets $U
\subseteq \Gamma^{\infty}$ with $\alpha \in U$ we have $U \cap L \neq
\emptyset$.
% It is given by
% $\overline{L} = \set{\alpha \in \Gamma^{\infty}}{\forall U \subseteq
%   \Gamma^{\infty}\text{ open with } \alpha \in U \colon U \cap L \neq
%   \emptyset}$. 
The \emph{interior} of $L$ is the union of all open sets contained in
$L$.  It can be constructed as the complement of the closure of its
complement.
For languages $L$ and $K$ we define the right quotient as a language
of finite words by $L/K = \set{u \in \Gamma^*}{u \alpha \in L \text{
    for some } \alpha \in K}$. 
In particular, we have $$L/A^\infty =
\set{u \in \Gamma^*}{u \alpha \in L \text{ for some } \alpha \in
   A^{\infty}}.$$

 The following
proposition gives a description of the closure in the alphabetic
topology in terms of arrow languages $\overrightarrow{W}$ plus
some alphabetic restrictions.

\begin{proposition}\label{prop:closure}
  In the alphabetic topology we have $\ov{A^{\im}} = \bigcup_{A
    \subseteq B} B^{\im}$ and
  \begin{equation*}
    \ov{L} = 
    \bigcup_{A
      \subseteq \Gamma} \left( \overrightarrow{L/A^\infty} \cap
      A^{\im} \right) = \bigcup_{A
      \subseteq \Gamma} \left( \overrightarrow{L/A^\infty} \cap
      \ov{A^{\im}} \right).
  \end{equation*}
\end{proposition}

\proof
It is elementary to show $\ov{A^{\im}} = \bigcup_{A \subseteq B}
B^{\im}$.
We first show %the inclusion 
$\ov{L} \subseteq \bigcup_{A \subseteq
  \Gamma} \big( \overrightarrow{L/A^\infty} \cap A^{\im} \big)$. Let
$\alpha \in \ov{L}$ with $\alpha \in A^{\im}$.  For all prefixes $u$
of $\alpha$ we find $v$ such that $\alpha \in u vA^{\infty}$. We have
$uv A^{\infty} \cap L \neq \emptyset$; and thus $uv \in
L/A^\infty$. This shows $\alpha \in \overrightarrow{L/A^\infty}$.
  
The inclusion $\bigcup_{A \subseteq \Gamma} \big(
\overrightarrow{L/A^\infty} \cap A^{\im} \big) \subseteq \bigcup_{A
  \subseteq \Gamma} \big( \overrightarrow{L/A^\infty} \cap
\ov{A^{\im}} \big)$ is trivial.

Let now $\alpha \in \overrightarrow{L/A^\infty} \cap B^{\im}$ with $A
\subseteq B$. Since $L/A^\infty \subseteq L/B^\infty$, we have $\alpha
\in \overrightarrow{L/B^\infty} \cap B^{\im}$. Let $u \in \Gamma^*$
with $\alpha = u \beta$ and $\beta \in B^{\infty}$. We have to show $u
B^{\infty} \cap L \neq \emptyset$. Since $\alpha \in
\overrightarrow{L/B^\infty}$ there is some $v \in \Gamma^*$ with $uv
\leq \alpha$ and $uv \in L/B^\infty$. This means $uv\gamma \in L$ for
some $\gamma \in B^{\infty}$. Since $\beta \in B^{\infty}$ we have $v
\in B^*$. Hence $v\gamma \in B^{\infty}$ and thus $uv\gamma \in u
B^{\infty} \cap L \neq \emptyset$ as desired.
\qed

The following corollary  generalizes a well-known fact for 
the Cantor topology to the (finer) alphabetic topology.
This result  will be used in Section~\ref{sec:delta2}.

\begin{corollary}\label{cor:closedisdet}
  Let $ L \subseteq \Gamma^{\infty}$ be a regular language. Then its
  closure in the alphabetic topology $\ov L$ is deterministic.
\end{corollary}

\proof 
Deterministic languages are closed under finite union and finite
intersection. For a letter $a$ the language $\ov{\{a\}^{\im}}$ is
deterministic as it is the language of words having infinitely many
$a$'s.  Hence $\ov{A^{\im}} = \bigcap_{a \in A}\ov{\{a\}^{\im}}$ is
deterministic, too. The result follows.
\qed

\begin{corollary}\label{cor:closure}
  Given a regular language $ L \subseteq \Gamma^{\infty}$, we can
  decide whether $L$ is closed (open resp., clopen resp.).
\end{corollary}

\proof 
We may assume that $L$ is specified by some NFA for $ L \cap \Gamma^*$
and by some B{\"u}chi automaton for $L \cap\Gamma^{\oo}$. The construction
of an NFA recognizing ${L/A^\infty}$ is standard. Since
${L/A^\infty}\subseteq \GA^*$ we can assume that the NFA is
deterministic, and we can view it as a (deterministic) Büchi automaton
recognizing $\overrightarrow{L/A^\infty}\cap \GA^\oo$. Intersection
with ${A^{\im}}$ yields a B{\"u}chi automaton for $\ov{L}\cap A^{\im}$
and $A\neq \emptyset$. Thus, we can test $\ov{L}\cap A^{\im} \subseteq
L$ for all $A$.  This implies that we can test $L = \ov{L}$. The
result for \emph{open} and \emph{clopen} follows since regular
languages are effectively closed under complementation.
\qed

Actually, we have a more precise statement than pure decidability.
In the following, $\mathrm{PSPACE}$ denotes as usual the 
class of problems which can be decided by some polynomially space bounded 
(deterministic) Turing machine. 

\begin{theorem}\label{thm:pspaceclosure}
  The following problem is $\mathrm{PSPACE}$-complete: \\
  Input: A Büchi automaton $\A$ with $L(\A)\subseteq \Gamma^{\omega}$. \\
  Question: Is the regular language $L(\A)$ closed?
\end{theorem}

\proof
We can check in $\mathrm{PSPACE}$ whether a regular language $L
\subseteq \Gamma^{\omega}$ is closed: Let $L = L(\mathcal{A})$ for
some non-deterministic B{\"u}chi automaton $\mathcal{A}$. We verify $L
= \overline{L}$ using the characterization of $\overline{L}$ given in
Proposition~\ref{prop:closure}. We can
check in $\mathrm{PSPACE}$ whether 
two B\"uchi automata are equivalent, see~\cite{svw87tcs}.
In particular, we can
check in $\mathrm{PSPACE}$ whether $L \cap A^{\im} =
\overrightarrow{L/A^{\infty}} \cap A^{\im}$  for all $A \subseteq \Gamma$.

It is $\mathrm{PSPACE}$-hard to decide whether a regular language $L
\subseteq \Gamma^{\omega}$ is closed: We use a reduction of the
problem whether $L(\mathcal{A}) = \Gamma^*$ for some NFA
$\mathcal{A}$, see~\cite{ms72focs}. We can assume that $1 \in
L(\mathcal{A})$.  Let $c \not\in \Gamma$ be a new letter. We can
construct a non-deterministic B{\"u}chi automaton $\mathcal{B}$ such
that $L(\mathcal{B}) = \set{w_1 c w_2 c \cdots \in (\Gamma \cup
  \smallset{c})^{\omega}} {\exists i \colon w_i \in
  L(\mathcal{A})}$. The closure of $L(\mathcal{B})$ is $K = \set{w_1 c
  w_2 c \cdots \in (\Gamma \cup \smallset{c})^{\omega}}{\forall i
  \colon w_i \in \Gamma^*} = (\Gamma^* c)^{\omega}$. Hence,
$L(\mathcal{A}) = \Gamma^*$ if and only if $L(\mathcal{B}) = K$ if and
only if $L(\mathcal{B})$ is closed.
\qed

According to Proposition~\ref{prop:closure} the alphabetic closure is
a union over languages of type $\overrightarrow{L/A^\infty}$ or
$\overrightarrow{L/A^\infty} \cap \ov{A^{\im}}$.  But these pieces do
not themselves need to be closed, as we can see in the following
example.
  
\begin{example}
  Let $A =
  \smallset{a}$, $B = \smallset{a,b}$, and $L = a^* (ab)^* b
  a^{\omega}$. Then $L/A^\infty = a^* (ab)^* b a^*$ and $L/B^\infty$
  is the set of all finite prefixes of words in $L$. We have
  $\overrightarrow{L/A^\infty} = a^* (ab)^* b a^{\infty}$ and
  $\overrightarrow{L/A^\infty} \cap \ov{A^{\im}} = a^* (ab)^* b
  a^{\omega} = L$. The language $\overrightarrow{L/A^\infty}$ is
  open but neither  $\overrightarrow{L/A^\infty}$ nor
  $\overrightarrow{L/A^\infty} \cap \overline{A^{\im}}$ is closed 
  in the alphabetic topology, because $(ab)^\omega$ belongs to both
  closures. We have $\overrightarrow{L/B^\infty} = a^* (ab)^* b
  a^{\infty} \cup a^* (ab)^{\omega}$ and $\overrightarrow{L/B^\infty}
  \cap B^{\im} = a^* (ab)^{\omega}$. Both sets are closed. Actually,
  $\ov{L} = L \cup a^* (ab)^{\omega}$ in the alphabetic topology.
  
  The alphabetic closure $\ov{L}$ is not closed in the Cantor topology
  since $a^{\omega} \not\in \ov{L}$, but every Cantor-open neighborhood 
  of $a^{\omega}$ contains a word $a^n (ab)^{\omega}$ for some 
  $n\in \N$. 
\end{example}

Frequently we apply the closure operator to polynomials.  A
\emph{polynomial} is a finite union of monomials.  A \emph{monomial}
(of \emph{degree} $k$) is a language of the form $A_1^* a_1 \cdots
A_k^* a_k A_{k+1}^{\infty}$ with $a_i \in \Gamma$ and $A_i \subseteq
\Gamma$.  In particular, $A_1^* a_1 \cdots A_k^* a_k$ is a monomial
with $ A_{k+1} = \emptyset$.  The set $A^*$ is a polynomial since
$A^*= \emptyset^\infty \cup \bigcup_{a \in A} A^* a$. It is not hard
to see that polynomials are closed under intersection.  Thus, $A_1^*
a_1 \cdots A_k^* a_k A_{k+1}^* = A_1^* a_1 \cdots A_k^* a_k
A_{k+1}^\infty \cap \Gamma^*$ is in our language a polynomial, but not
a monomial unless $A_{k+1} = \emptyset$.  

A monomial $P = A_1^* a_1
\cdots A_k^* a_k A_{k+1}^{\infty}$ is called \emph{unambiguous}, if for every
$\alpha \in P$ there exists a unique factorization $\alpha = u_1 a_1
\cdots u_k a_k \beta$ such that $u_i \in A_i^*$ and $\beta \in
A_{k+1}^{\infty}$. A polynomial is called \emph{unambiguous}, if it is a
finite union of unambiguous monomials.

\begin{example}\label{ex:upol}
  For $\Gamma = \oneset{a,b}$ the language $\Gamma^* ab \Gamma^\infty$
  can be written as an unambiguous monomial, because:
  \begin{equation*}
    \Gamma^* ab \Gamma^\infty =b^*\, a\,  a^*\, b\, 
    \smallset{a,b}^\infty .
  \end{equation*}
  Similarly, $\Gamma^* ab \Gamma^*$ can be written as an unambiguous
  polynomial.  However, for $\Gamma = \oneset{a,b,c}$ the situation is
  different. Neither $\Gamma^*ab\,\Gamma^*$ nor
  $\Gamma^*ab\,\Gamma^\infty$ is unambiguous.  Their syntactic monoid
  is the monoid $M = \oneset{1,a,b,c,ba,0}$ defined in
  Example~\ref{ex:damon}, which is not in $\DA$ as shown there. So the
  claim follows by Theorem~\ref{thm:fo2}.
\end{example}

It follows from the definition of the alphabetic topology that
polynomials are open.  Actually, it is the coarsest topology with this
property. The crucial observation is that we have a syntactic
description of the closure of a polynomial as a finite union of other
polynomials. For later use we make a more precise statement by
considering the closure with respect to different subsets $B$ at
infinity.

\begin{lemma}\label{lem:polclosure}
  Let $P = A_1^* a_1 \cdots A_k^* a_k A_{k+1}^{\infty}$ be a monomial
  and $L = P \cap B^{\im}$ for some $B \subseteq A_{k+1}$. Then the
  closure of $L$ in the alphabetic topology is given by
  \begin{equation*}
    \overline{L} = 
    \bigcup_{\smallset{a_i,\ldots,a_k}\cup B \subseteq A\subseteq A_i}
    A_1^* a_1 \cdots A_{i-1}^* a_{i-1} A_i^{\infty} \cap A^{\im}.
  \end{equation*}
\end{lemma}

\proof
First consider an index $i$ with $1 \leq i \leq k+1$ such that
$\oneset{a_i, \ldots, a_k}\cup B \subseteq A \subseteq A_i$.  Let
$\alpha\in A_1^* a_1 \cdots A_{i-1}^* a_{i-1} A_i^{\infty} \cap
A^{\im}$. We have to show that $\alpha$ is in the closure of $L$.  Let
$\alpha = u \beta$ with $u \in A_1^* a_1 \cdots A_{i-1}^* a_{i-1}
A_i^*$ and $\beta \in \cpx{A}$. We show that $uA^\infty \cap L \neq
\emptyset$. Choose some $\gamma \in \cpx{B}$. As $B \subseteq A_{k+1}$
holds by hypothesis, we see that $ua_{i} \cdots a_k \gamma \in P$, and
hence $ua_{i} \cdots a_k \gamma \in uA^\infty \cap L$.

Let now $\alpha \in \overline{L}$ and write $\alpha \in u v_1 \cdots
v_{k+1} \cpx{A}$ with $\alp(v_j) = A$. There exists $\gamma \in
A^{\infty}$ such that $uv_1 \cdots v_{k+1}\gamma \in P\cap B^{\im}$.
This implies $B \subseteq A$.  Since $uv_1 \cdots v_{k+1}\gamma \in
A_1^* a_1 \cdots A_{k}^* a_{k} A_{k+1}^\infty$ there are some $1 \leq
i, j \leq k+1$ such that $ uv_1 \cdots v_{j-1}$ belongs to $A_1^* a_1
\cdots A_{i-1}^* a_{i-1} A_i^*$, $v_j\in A_i^*$, and
$v_{j+1} \cdots v_{k+1}\gamma \in A_i^* a_i \cdots A_{k}^* a_{k}
A_{k+1}^\infty\cap A^\infty$.  Therefore $\oneset{a_i, \ldots, a_k}
\subseteq A \subseteq A_i$, too.  It follows that $\alpha \in A_1^*
a_1 \cdots A_{i-1}^* a_{i-1} A_i^{\infty} \cap A^{\im}$.
\qed

\begin{example}\label{ex:clos}
Let $\Gamma = \oneset{a,b,c}$ and  $L = \Gamma^*ab\,\Gamma^*$.
Its closure %in the alphabetic topology 
is given by
  \begin{equation*}
    \overline{L} = 
     \Gamma^*ab\,\Gamma^\infty  \cup \oneset{a,b}^{\im} \cup \oneset{a,b,c}^{\im}
     = \Gamma^*ab\,\Gamma^\infty \cup \Gamma^{\im}. 
  \end{equation*}
\end{example}

As usual, let $L\subseteq \GA^\infty$ be a regular
language. Let us define $ tf^\oo \leq_L se^\oo $ for linked
pairs $(s,e)$, $(t,f)$ by the implication:
\begin{equation*}
  [s][e]^\oo \subseteq L \implies  [t][f]^\oo \subseteq L.
\end{equation*}
With this notation we can give an algebraic characterization of being
open.

\begin{lemma}\label{lem:algopen}
  A regular language $L \subseteq \Gamma^\infty$ is open in the
  alphabetic topology if and only if for all linked pairs $(s,e)$,
  $(t,f)$ of $M = \Synt(L)$ with $t,f \in M_e$ we have $stf^\oo
  \leq_L se^\oo$.
\end{lemma}

\proof 
Let $L$ be open and $\alpha \in [s][e]^\oo \subseteq L$. We find a
finite prefix $u\in [s]$ of $\alpha$ such that $\alpha \in u A^\infty
\subseteq L$. Since $t,f \in M_e$ we may assume $\alp(vw) \subseteq A$
for some $v \in [t]$ and $w \in [f]$. Hence, $uvw^\omega \in
[st][f]^\omega \subseteq L$. This shows $stf^\oo \leq_L se^\oo$.
 
For the converse, suppose that for all linked pairs $(s,e), (t,f)$ of
$M = \Synt(L)$ with $t,f \in M_e$ we have $stf^\oo \leq_L se^\oo.$ Let
$\alpha \in [s][e]^\oo \subseteq L$. Write $\alpha = u\beta$ with $u
\in [s]$ and $\beta \in [e]^\oo \cap \cpx{A}$.  Now, any $\gamma \in
A^\infty$ can be written as $\gamma \in [t][f]^\oo$ for some linked
pair with $t,f \in M_e$. Indeed, we have $A^* \subseteq [M_e]$:
consider $a \in A$ and let $p,q \in A^*$ such that $paq \in [e]$. Then
$a \in [M_e]$ and therefore $A \subseteq [M_e]$. Since $M_e$ is a
submonoid, $[M_e]$ is a submonoid of $\Gamma^*$ and hence $A^*
\subseteq [M_e]$. By assumption $u\gamma \in [st][f]^\oo \subseteq
L$. It follows $u A^\infty \subseteq L$, i.e., $L$ is open.
\qed

%%%%%%%%%%%%%%%%%%%%%%%%%%%%%%%%%%%%%%%%%%%%%%%%%%%%%
\section{The fragment $\mathbf{\Sigma_2}$}\label{sec:sigma2}

By a (slight extension of a) result of Thomas~\cite{tho82} on
$\oo$-languages we know that a language $L\subseteq \Gamma^\infty$ is
definable in $\Sigma_2$ if and only if $L$ is a polynomial. However,
this statement alone does not yield decidability.  It turns out
that we obtain decidability by a combination of an algebraic and a
topological criterion. (This decidability result has also been shown 
independently by
Boja{\'n}czyk~\cite{boj08fossacs} using different techniques.)  We know
that polynomials are open.  Therefore, we concentrate on algebra.

\begin{lemma}\label{lem:Pol:LocHigh}
  If $L \subseteq \Gamma^{\infty}$ is a polynomial, then all
  idempotents of $\Synt(L)$ are locally top.
\end{lemma}

\proof
By $h_L$ we denote the syntactic homomorphism $\Gamma^* \to \Synt(L)$.
Let $n\in \N$ such that $L$ is a finite union of monomials of degree
less than $n$.  Let $h_L(e)$ be idempotent; in particular $e^n \equiv_L
e$.  For $e \equiv_L f$ we may assume that $\alp(f) \subseteq
\alp(e)$. This means we take the maximal possible alphabet for
$e$. Let $s \in \alp(e)^*$. We want to show that $xeseyz^\omega
\in L$ if $xeyz^\omega \in L$.
  
Suppose $u = xe^nyz^{\omega} \in A_1^* a_1 \cdots A_k^* a_k
A_{k+1}^{\infty} \subseteq L$ and $k<n$. Since there are at most $n-1$
letters $a_i$, some factor $e$ of $u$ lies completely within one of
the $A_i^*$ or within $A_{k+1}^{\infty}$, i.e., $\alp(e) \subseteq
A_i$ for some $1 \leq i \leq k+1$. Hence, $ese \in A_i^*$ and
$xe^{n_1}se^{n_2}yz^{\omega} \in A_1^* a_1 \cdots A_k^* a_k
A_{k+1}^{\infty} \subseteq L$ for some $n_1,n_2 \geq 1$. Since $h_L(e)$
is idempotent, it follows that $xeyz^{\omega} \in L$ implies
$xeseyz^{\omega} \in L$. Similarly, $x(ey)^\omega \in L$ implies
$x(esey)^\omega \in L$ and therefore $ese \leq_L e$ for all $s \in
\alp(e)^*$, i.e., $h_L(e)$ is locally top.
\qed

\begin{theorem}\label{thm:sigma2}
  Let $L \subseteq \Gamma^{\infty}$ be a regular language. The
  following five assertions are equivalent:
  \begin{enumerate}
  \item\label{aaa:sigma2} $L$ is $\Sigma_2$-definable.
  \item\label{bbb:sigma2} $L$ is a polynomial.
  \item\label{ccc:sigma2} $L$ is open in the alphabetic topology and
    all idempotents of $\Synt(L)$ are locally top.
  \item\label{eee:sigma2} The syntactic monoid $M = \Synt(L)$ and the
    syntactic order $\leq_L$ satisfy:
    \begin{enumerate}
    \item\label{eee0:sigma2} For all linked pairs $(s,e)$, $(t,f)$
      with $t,f \in M_e$ we have $stf^\oo \leq_L se^\oo$.
    \item\label{eee1:sigma2} $e=e^2$ and $s \in M_e$ implies $ese
      \leq_L e$.
    \end{enumerate}
  \item\label{ddd:sigma2} The following three conditions hold for some
    homomorphism $h : \Gamma^* \to M$ which weakly recognizes $L$:
    \begin{enumerate}
    \item\label{ddd0:sigma2} $L$ is open in the alphabetic topology.
    \item\label{ddd1:sigma2} All idempotents of $M$ are locally top.
    \item\label{ddd2:sigma2} $L$ is downward closed (on finite
      prefixes) for $h$.
    \end{enumerate}
  \end{enumerate}  
\end{theorem}

\proof
  ``\ref{aaa:sigma2}~$\Leftrightarrow$~\ref{bbb:sigma2}'': This is a
  slight modification of a result by Thomas~\cite{tho82}.

  ``\ref{bbb:sigma2}~$\Rightarrow$~\ref{ccc:sigma2}'': By definition,
  polynomials are open in the alphabetic topology. In
  Lemma~\ref{lem:Pol:LocHigh} it has been shown that all idempotent
  elements are locally top.

  ``\ref{ccc:sigma2}~$\Leftrightarrow$~\ref{eee:sigma2}'': The
  equivalence of $L$ being open and ``\ref{eee0:sigma2}'' is
  Lemma~\ref{lem:algopen}. Property ``\ref{eee1:sigma2}'' is the
  definition of all elements being locally top.

  ``\ref{eee:sigma2}~$\Rightarrow$~\ref{ddd:sigma2}'': Let $h = h_L$
  be the syntactic homomorphism onto the syntactic monoid $M=
  \Synt(L)$. Since $L$ is regular, the homomorphism $h$ strongly
  recognizes $L$. Applying Lemma~\ref{lem:algopen}, the property
  ``\ref{ddd0:sigma2}'' follows from ``\ref{eee0:sigma2}'' and
  ``\ref{ddd1:sigma2}'' trivially follows from ``\ref{eee1:sigma2}''.
  The condition ``\ref{ddd2:sigma2}'' holds for $\Synt(L)$ by
  Lemma~\ref{lem:SyntL:DownClosed}.
  
  ``\ref{ddd:sigma2}~$\Rightarrow$~\ref{bbb:sigma2}'': Consider
  $\alpha \in L$ with $\im(\alpha) = A$. By ``\ref{ddd0:sigma2}'' the
  language $L$ is open. Hence, there exists a prefix $u$ of $\alpha$
  such that $\alpha \in u A^{\infty} \subseteq L$. From the case of
  finite words and the hypothesis ``\ref{ddd1:sigma2}'' on $M$, we
  know that $P = \set{v \in \Gamma^*}{h(v) \leq h(u)}$ is a
  polynomial. We can assume that all monomials in $P$ end with a
  letter.  We define the polynomial $P_{\alpha} =
  PA^{\infty}$. Clearly, $L \subseteq \bigcup \set{P_{\alpha}}{\alpha
    \in L}$ and this union is finite since $M$ is finite. It remains
  to show that $P_{\alpha} \subseteq L$ for $\alpha \in L$. Let $v \in
  P$ and $\beta \in A^{\infty}$. We know $u\beta \in L$ and there
  exists a linked pair $(s,e)$ such that $u\beta \in [s][e]^{\omega}
  \subseteq L$.  Now, there exists $w \gamma = \beta$ such that $uw
  \in [s]$ and $\gamma \in [e]^{\omega}$. By definition of $P$, we
  have $h(v) \leq h(u)$ and therefore $t = h(vw) \leq h(uw) = s$. It
  follows $v \beta = v w \gamma \in [t][e]^{\omega} \subseteq L$ by
  ``\ref{ddd2:sigma2}''. This shows $P_{\alpha} \subseteq L$ and thus
  $L = \bigcup \set{P_{\alpha}}{\alpha \in L}$.
\qed

\begin{corollary}\label{cor:DecideSigma2}
  It is decidable whether a regular language is
  $\Sigma_2$-definable.
\end{corollary}

\proof
The syntactic congruence is computable and the conditions in
``\ref{ccc:sigma2}'' (or ``\ref{eee:sigma2}'') of
Theorem~\ref{thm:sigma2} are decidable.
\qed

\begin{remark}\label{rem:njal}
  An $\oo$-language $L \subseteq \GA^\oo$ is
  $\Sigma_2$-definable, if $L = \set{\alpha \in \GA^\oo}{\alpha
    \models \phi}$ for some $\phi \in \Sigma_2$.  This is equivalent with
  $L \cup \GA^*$ being $\Sigma_2$-definable as a subset of $\GA^\infty$.
  Thus, the decidability of Corollary~\ref{cor:DecideSigma2} transfers
  to $\oo$-regular languages.
\end{remark}

Of course, complementation yields dual results for the fragment
$\Pi_2$. In particular, $\Pi_2$-definable languages are closed
in the alphabetic topology.

%%%%%%%%%%%%%%%%%%%%%%%%%%%%%%%%%%%%%%%%%%%%%%%%%%%%%

\section{Two variable first-order logic}\label{sec:fo2}

Etessami, Vardi, and Wilke have given a characterization of $\FO^2$ in
terms of unary temporal logic~\cite{evw02}.  In the same paper, they
considered the satisfiability problem for $\FO^2$. We continue the
study of $\FO^2$ over infinite words. It will turn our that the
fragments $\FO^2$ and $\Sigma_2$ are incomparable. Therefore, it makes
sense to also consider $\FO^2 \cap \Sigma_2$ and $\FO^2 \cap \Pi_2$.

\subsection[The fragment $\FO^2$ and the strict alphabetic
  topology]{The fragment $\mathbf{\BFO^2}$ and the strict alphabetic
  topology}\label{sec:fo2da}

%\subsection{$\FO^2 = \DA$ and the strict alphabetic topology}\label{sec:fo2da}

This section yields the algebraic characterization of $\FO^2$
in terms of the variety $\DA$.
The following lemma can be proved essentially in the same way as for
finite words. The result is also (implicitly) stated in the
habilitation thesis of Wilke~\cite{wil98}.

\begin{lemma}\label{lem:FO2:DA}
  Let $L \subseteq \Gamma^{\infty}$ be $\FO^2$-definable. Then the
  syntactic monoid $\Synt(L)$ is in $\DA$.
\end{lemma}

\proof
Let $L = L(\varphi)$ for some $\FO^2$-formula of quantifier depth $n$.
Let $e^2 = e \in M = \Synt(L)$ and let $s \in M_e$. We can choose
words $v,w \in \Gamma^*$ such that $h_L(v) = s$, $h_L(w) = e$, and,
moreover, $\alp(v) \subseteq \alp(w)$. Now, consider words of the form
$\alpha = x w^n v w^n y z^{\omega}$, $\alpha' = x w^n y z^{\omega}$
and $\beta = x (w^n v w^n y)^{\omega}$, $\beta' = x (w^n
y)^{\omega}$. It is easy to show that the second player has a winning
strategy in the $n$-round \EF{} game for $\FO^2$ on $(\alpha,\alpha')$
and also on $(\beta,\beta')$. A description of the game can be found
in~\cite{imm82} and the winning strategy is a modification of the
proof in the finitary case~\cite{tw98stoc}. The game equivalence
implies that both words in each pair satisfy the same $\FO^2$
sentences of quantifier depth no more than $n$. In particular, $\alpha
\in L$ if and only if $\alpha' \in L$. Analogously, $\beta \in L$ if
and only if $\beta' \in L$.  Thus, $\Synt(L) \in \DA$.
\qed

A set like $A^{\im}$ is $\FO^2$-definable, but it is neither open
nor closed in the alphabetic topology, in general.  Therefore, we need
a refinement of the alphabetic topology. As a basis for the
\emph{strict alphabetic topology} we take all sets of the form $u
\cpx{A}$. Thus, more sets are open (and closed) than in the alphabetic
topology. Another way to define the strict alphabetic topology is to
say that it is the coarsest topology on $\Gamma^\infty$ where all sets
of the form $A_1^* a_1 \cdots A_k^* a_k A_{k+1}^{\infty}\cap B^{\im}$
are open.  The strict alphabetic topology is not used outside this
section, but it is essential here in order to prove the converse of
Lemma~\ref{lem:FO2:DA}.

\begin{lemma}\label{lem:DA:clopen}
  If $L \subseteq \Gamma^{\infty}$ is strongly recognized by some
  homomorphism $h : \Gamma^* \to M \in \DA$, then $L$ is clopen in the
  strict alphabetic topology.
\end{lemma}

\proof
Since $h$ strongly recognizes $\Gamma^{\infty} \setminus L$ as
well, it is enough to show that $L$ is open. Let $\alpha \in L$ with
$\alpha \in [s][e]^{\omega}$ for some linked pair $(s,e)$ and let $A =
\im(\alpha)$. We show that $[s] \cpx{A} \subseteq L$. Indeed, let
$\beta \in [s] \cpx{A}$. Then we have $\beta = uv\gamma$ with $h(u) =
s$, $h(v) = r$, $\gamma \in [f]^{\omega}$ where $v \in A^*$,
$\alp(\gamma) = \im(\gamma) = A$, and $(r,f)$ is a linked pair. Since
$M \in \DA$, we obtain $s = se = serfe = srfe$ and $efe = e$ and $fef
= f$. We have $[sr][fef]^{\omega}
  \cap [srfe][efe]^{\omega} \neq \emptyset$ and 
$[srfe][efe]^{\omega} = [s][e]^{\omega} \subseteq L$. 
Since $h$ strongly recognizes $L$, we have
$[sr][f]^{\omega} = [sr][fef]^{\omega}
   \subseteq L$, too. In particular, $\beta \in L$.
\qed

\begin{lemma}\label{lem:DAclosed:UPol}
  If $L$ is closed in the strict alphabetic topology and if $L$ is
  weakly recognized by some homomorphism $h : \Gamma^* \to M \in \DA$,
  then $L$ is a finite union of languages $A_1^* a_1 \cdots A_k^* a_k
  A_{k+1}^{\infty} \cap A_{k+1}^{\im}$, where each $A_1^* a_1 \cdots
  A_k^* a_k A_{k+1}^{\infty}$ is an unambiguous monomial.
\end{lemma}

\proof
Let $\alpha \in L$. Write $\alpha = u \beta$ with $\beta \in \cpx{A}$
for some $A \subseteq \Gamma$. There is a linked pair $(s,e)$ with
$\alpha \in [s][e]^{\omega} \subseteq L$ and we may assume $h(u) = s$
and $\beta \in [e]^{\omega}$. For $A = \emptyset$ we have $[s]
\subseteq L$ and, using our knowledge about the finite case, we may
include $[s]$ in our finite union of unambiguous
polynomials. Therefore, let $A \neq \emptyset$. We may choose an
unambiguous monomial $P = A_1^* a_1 \cdots A_k^* a_k \subseteq [s]$
such that $u \in P$ and each last position of every letter $a \in
\oneset{a_1, \ldots, a_k} \cup A_1 \cup \cdots \cup A_k$ occurs
explicitly as some $a_j$ in the expression $P$. Note that $[s]$ is a
finite union of such monomials. Moreover, we may assume that $uv \in
P$ for infinitely many prefixes $v \leq \beta$. Each such $uv$ can
uniquely be written as $uv = v_1 a_1 \cdots v_k a_k$ with $v_i \in
A_i^*$. This yields a vector in $\N^k$ by $(\abs{v_1 a_1},\abs{v_1 a_1
  v_2 a_2}, \ldots, \abs{v_1 a_1 \cdots v_k a_k})$ for every $uv \in
P$.  By Dickson's Lemma~\cite{dickson1913}, every infinite sequence in
$\N^k$ contains an infinite subsequence which is non-decreasing in
every component. Therefore, we may assume that the sequence of vectors
induced by the prefixes $uv$ is in no component decreasing when $uv$
gets longer. In addition (after removing finitely many $uv$'s) we may
assume there is some $i \geq 0$ such that the component $\abs{v_1 a_1
  \cdots v_i a_i}$ is constant and $\abs{v_1 a_1 \cdots v_i a_i
  v_{i+1} a_{i+1}}$ is strictly increasing. It follows that we may
assume $\oneset{a_{i+1}, \ldots, a_k} \subseteq \alp(v_{i+1}) = A
\subseteq A_{i+1}$. In particular, $\alpha \in A_1^* a_1 \cdots A_i^*
a_i \cpx{A}$. It is clear that this expression is unambiguous.

It remains to show $A_1^* a_1 \cdots A_i^* a_i \cpx{A} \subseteq
L$. Consider $u' \gamma$ with $u' \in A_1^* a_1 \cdots A_i^* a_i$ and
$\gamma \in \cpx{A}$. Since $L$ is closed, it is enough to show that
$u' \gamma$ belongs to the closure of $L$ in the strict alphabetic
topology. Choose any prefix $w \leq \gamma$.  It is enough to show
that $u' w \cpx{A} \cap L \neq \emptyset$. Let $z \in \Gamma^*$ with
$\alp(z) = A$ and $h(z) = e$. Since $w \in A^* \subseteq A_{i+1}^*$,
we have $u' w a_{i+1} \cdots a_k \in P \subseteq [s]$. Hence $u' w
a_{i+1} \cdots a_k z^{\omega} \in [s][e]^{\omega} \subseteq L$.
\qed

The next statement follows again as in the case of finite words.

\begin{lemma}\label{lem:UPol:FO2}
  Every language $A^{\im}$ and every unambiguous monomial $A_1^* a_1
  \cdots A_k^* a_k A_{k+1}^{\infty}$ is $\FO^2$-definable.
\end{lemma}

\proof
The language of non-empty words in $A^{\im}$ is defined by the
$\FO^2$-sentence
\begin{equation*}
  \bigwedge_{a \in A} \Big( \forall x \exists y \colon x < y \,\wedge\, 
  \lambda(y) = a\Big) \ \wedge \ 
  \bigwedge_{b \not\in A} \Big( \exists x \forall y \colon x < y \,\wedge\,
  \lambda(y) \neq b\Big).
\end{equation*}
We use induction on $k$ in order to show that $P = A_1^* a_1 \cdots
A_k^* a_k A_{k+1}^{\infty}$ is $\FO^2$-definable. Clearly, for $k = 0$
this is true. Let now $k \geq 1$. By unambiguity, we cannot have
$\oneset{a_1, \ldots, a_k} \subseteq A_1 \cap A_{k+1}$ since for $(a_1
\cdots a_k)^2$ there would exist two different factorizations. First,
suppose $a_i \not\in A_{k+1}$. Let $\alpha = \alpha_1 a_i \alpha_2 \in
P$ where $a_i \not\in \alp(\alpha_2)$. There are two possibilities:
the last $a_i$ of $\alpha$ could be one of the $a_j$'s, $i \leq j \leq
k$, and then
\begin{equation*}
  \alpha_1 \in A_1^* a_1 \cdots A_j^*, \quad
  a_i = a_j, \quad
  \alpha_2 \in A_{j+1}^* a_{j+1} \cdots A_k^* a_k A_{k+1}^{\infty}
\end{equation*}
or it matches some $A_j^*$, $i < j < k+1$ and then
\begin{equation*}
  \alpha_1 \in A_1^* a_1 \cdots A_j^*, \quad
  a_i \in A_j, \quad
  \alpha_2 \in A_{j}^* a_{j} \cdots A_k^* a_k A_{k+1}^{\infty}.
\end{equation*}
In any case, the remaining four polynomials are unambiguous and their
degree is strictly smaller than $k$. Hence, by induction we have
$\FO^2$-formulas describing them. Obviously, we can also express
intersections with languages of the form $B^*$ or $B^{\infty}$ for $B
\subseteq \Gamma$. So there is a finite list of $\FO^2$-formulas such
that for each $\alpha \in P$ there are formulas $\varphi$ and $\psi$
from the list and a letter $a \in \Gamma$ with $\alpha \in L(\varphi)
a L(\psi) \subseteq P$ and $L(\psi) \subseteq (\Gamma \setminus
\smallset{a})^{\infty}$. Now, the last $a$-position $x$ in every
$\alpha \in L(\varphi) a L(\psi)$ is uniquely defined by
\begin{equation*}
  \xi(x) \ \ = \ \ \lambda(x) = a \,\wedge\, 
  \forall y\colon x<y \Rightarrow \lambda(y) \neq a.
\end{equation*}
Using relativization techniques, we now define $\FO^2$-sentences
$\varphi_{<a}$ and $\psi_{>a}$ such that $L(\varphi) a L(\psi) =
L\big(\varphi_{<a} \wedge \exists x \colon \xi(x) \wedge
\psi_{>a}\big)$. We give the inductive construction for $\psi_{>a}$. The
other one for $\varphi_{<a}$ is symmetric. Atomic formulas are
unchanged and Boolean connectives are straightforward. Existential
quantification is as follows:
$(\exists x\colon \zeta)_{>a} \ = \ \exists x \colon 
(\exists y\colon y<x \wedge \xi(y)) \wedge \zeta_{>a}$.

The case $a_i \not\in A_1$ is similar (using a factorization of
$\alpha$ at the first $a_i$-position).
\qed

\begin{theorem}\label{thm:fo2}
  Let $L \subseteq \Gamma^{\infty}$. The
  following assertions are equivalent:
  \begin{enumerate}
  \item\label{aaa:fo2} $L$ is $\FO^2$-definable.
  \item\label{bbb:fo2} $L$ is regular and $\Synt(L) \in \DA$.
  \item\label{ccc:fo2} $L$ is strongly recognized by some homomorphism
    $h : \Gamma^* \to M \in \DA$.
  \item\label{ddd:fo2} $L$ is closed in the strict alphabetic topology
    and $L$ is weakly recognized by some homomorphism $h : \Gamma^*
    \to M \in \DA$.
  \item\label{eee:fo2} $L$ is a finite union of sets of the form
    $A_1^* a_1 \cdots A_k^* a_k A_{k+1}^{\infty} \cap A_{k+1}^{\im}$,
    where each language $A_1^* a_1 \cdots A_k^* a_k A_{k+1}^{\infty}$ is an
    unambiguous monomial.
  \end{enumerate}
\end{theorem}

\proof
  ``\ref{aaa:fo2}~$\Rightarrow$~\ref{bbb:fo2}'': First-order definable
  languages are regular; $\Synt(L) \in \DA$ by Lemma~\ref{lem:FO2:DA}.
%
%  \noindent
  ``\ref{bbb:fo2}~$\Rightarrow$~\ref{ccc:fo2}'':
  Trivial, since $\Synt(L)$ strongly recognizes $L$.
%
%  \noindent
  ``\ref{ccc:fo2}~$\Rightarrow$~\ref{ddd:fo2}'': Strong recognition
  implies weak recognition; closure in the strict alphabetic topology
  follows by Lemma~\ref{lem:DA:clopen}.
%
%  \noindent
  ``\ref{ddd:fo2}~$\Rightarrow$~\ref{eee:fo2}'':
  Lemma~\ref{lem:DAclosed:UPol}.
%
%  \noindent
  ``\ref{eee:fo2}~$\Rightarrow$~\ref{aaa:fo2}'': Lemma~\ref{lem:UPol:FO2}.
\qed

Restricted to languages in $\Gamma^{*}$ the fragment $\FO^2$ is equal
to $\Delta_2$, hence it is equal to a fragment of $\Sigma_2$. In
general we have the following upper bound for languages over finite
and infinite words.

\begin{corollary}\label{cor:fo2:BS2}
  For languages in $\Gamma^{\infty}$ the fragment $\FO^2$ is contained
  in the Boolean closure of $\Sigma_2$.
\end{corollary}

\proof
Every $\FO^2$-definable language is a Boolean combination of
unambiguous monomials and alphabetic restrictions of the form
$A^{\im}$. By Theorem~\ref{thm:sigma2}, monomials are
$\Sigma_2$-definable and in the proof of Lemma~\ref{lem:UPol:FO2} we
have seen that $A^{\im}$ is definable in the Boolean closure of the
fragment $\Sigma_2$.
\qed

Recall that if a language $L \subseteq \Gamma^{\infty}$ is weakly
recognizable by some finite monoid, then it is also strongly
recognizable by a finite monoid. The same holds for aperiodic monoids:
if $L$ is weakly recognizable by some finite aperiodic monoid, then
there is a finite aperiodic monoid which strongly recognizes $L$.
Theorem~\ref{thm:fo2} suggests that this fails for $\DA$. Indeed, we
have the following example.

\begin{example}\label{ex:nwds}
  Let $\Gamma= \smallset{a,b,c}$.  Consider the congruence of finite
  index such that each class $[u]$ is defined by the set of words $v$
  where $u$ and $v$ agree on all suffixes of length at most 2. The
  quotient monoid of $\Gamma^*$ by this congruence is in $\DA$. In
  fact, it is a very simple monoid within $\DA$ since it is
  $\gL$-trivial (where $\gL$ is one of Green's relations, see
  e.g.~\cite{pp04}). Let $L = [ab]^\oo = (\Gamma^*ab)^\oo$. Then, by
  definition, $L$ is weakly recognizable in $\DA$; and $L$ is the
  language of all $\alpha$ which contain infinitely many factors of
  the form $ab$. This language is however not open in the strict
  alphabetic topology since $(cab)^\oo\in (\Gamma^*ab)^\oo$, but
  $(cab)^m(acb)^\oo\notin (\Gamma^*ab)^\oo$ for all $m \geq 0$.
\end{example}

%%%%%%%%%%%%%%%%%%%%%%%%%%%%%%%%%%%%%%%%%%%%%%%%%%%%%%%%%%%%%%%%%

\subsection[Unambiguous polynomials and the fragment $\FO^2 \cap
\Sigma_2$]{Unambiguous polynomials and the fragment $\mathbf{\BFO^2
    \cap \Sigma_2}$}\label{sec:upol}

In this section, we show that the intersection of $\FO^2$ and
$\Sigma_2$ has very natural descriptions involving topological notions
or unambiguous polynomials.

\begin{theorem}\label{thm:fo2si2}
  Let $L \subseteq \Gamma^{\infty}$. The following assertions are
  equivalent:
  \begin{enumerate}
  \item\label{aaa:fo2si2} $L$ is both $\FO^2$-definable and
    $\Sigma_2$-definable.
  \item\label{bbb:fo2si2} $L$ is $\FO^2$-definable and open in
    the alphabetic topology.
  \item\label{ccc:fo2si2} $L$ is an unambiguous polynomial,
    i.e., $L$ is a finite union of unambiguous monomials of the form
    $A_1^* a_1 \cdots A_k^* a_k A_{k+1}^{\infty}$.
  \item\label{ddd:fo2si2} $L$ is the interior  in
    the alphabetic topology of some
    $\FO^2$-definable language.
  \end{enumerate}
\end{theorem}

\proof
``\ref{aaa:fo2si2}~$\Rightarrow$~\ref{bbb:fo2si2}'':
Theorem~\ref{thm:sigma2}.

``\ref{bbb:fo2si2}~$\Rightarrow$~\ref{ccc:fo2si2}'': Let
$\alpha \in L \in \FO^2 \cap \Sigma_2$. By Theorem~\ref{thm:fo2} we
choose an unambiguous monomial $P = A_1^* a_1 \cdots A_k^* a_k$ (from
a given finite set depending on $L$) and $A \subseteq \Gamma$ such
that $P \cpx{A}$ is unambiguous and $\alpha \in P \cpx{A} \subseteq
L$. W.l.o.g.\ $A \neq \emptyset$. Let $A = \oneset{b_1, \ldots, b_m}$
and $B_i = A \setminus \smallset{b_i}$ and $R = B_1^* b_1 \cdots B_m^*
b_m$. Let $L$ be strongly recognized by $h : \Gamma^* \to M$. To every
sequence $v_1 \cdots v_n$ with $v_i \in \Gamma^*$ we can assign a
complete graph with vertices $\oneset{0,\ldots,n}$ where the edge
$(i,j)$ with $i<j$ is colored by the monoid element $h(v_{i+1} \cdots
v_j) \in M$.  By Ramsey's Theorem~\cite{ram30} there exists $r
\in \N$ such that for every sequence $v_1 \cdots v_r$ with $v_i \in
\Gamma^*$ there are $1 \leq j \leq \ell \leq r$ with $h(v_j \cdots
v_{\ell}) = e = e^2$ in $M$.

Trivially, we have $\alpha \in P R^r A^{\infty}$.  The monomial
$P R^r A^{\infty}$ is unambiguous and for some fixed language $L$ we
consider only finitely many of them.  We claim that $P R^r A^{\infty}
\subseteq L$. Let $\beta \in P R^r A^{\infty}$ and write $\beta = u
v_1 \cdots v_r \gamma$ with $u \in P$, $v_i \in R$, and $\gamma \in
A^{\infty}$. Choose $v_{j} \cdots v_{\ell} = v$ such that $h(v)$ is
idempotent. Then $u v_1 \cdots v_{\ell} v^{\oo} \in P \cpx{A}
\subseteq L$. Since $L$ is open and $\alp(v) = A$ we have $u v_1
\cdots v_{\ell} v^s A^{\infty} \subseteq L$ for some $s \in \N$. By
strong recognition and by idempotency of $h(v)$ we see that $\beta \in
u v_1 \cdots v_{\ell} A^{\infty} \subseteq L$. Therefore, $P R^r
A^{\infty} \subseteq L$.

``\ref{ccc:fo2si2}~$\Rightarrow$~\ref{aaa:fo2si2}'':
Theorem~\ref{thm:sigma2} and Theorem~\ref{thm:fo2}.
  
``\ref{ccc:fo2si2}~$\Rightarrow$~\ref{ddd:fo2si2}'': Trivial. 
 
``\ref{ddd:fo2si2}~$\Rightarrow$~\ref{bbb:fo2si2}'': It suffices to
show that the interior of an $\FO^2$-definable language is again
$\FO^2$-definable. Since $\FO^2$ is closed under complement, this is
equivalent to saying that the closure $\overline{K}$ of an
$\FO^2$-definable language $K$ is $\FO^2$-definable. By
Theorem~\ref{thm:fo2} we may assume that $K = P \cap B^{\im}$ where $P
= A_1^* a_1 \cdots A_k^* a_k A_{k+1}^{\infty}$ is an unambiguous
monomial and $B = A_{k+1}$.  By Lemma~\ref{lem:polclosure} we obtain
\begin{equation*}
  \overline{K} 
  = \bigcup_{\smallset{a_i,\ldots,a_k}\cup B \subseteq A\subseteq A_i} 
  \!\!\!\!\!\!\!\!
  A_1^* a_1 \cdots A_{i-1}^* a_{i-1} A_i^{\infty} \cap A^{\im}.
\end{equation*}
By Theorem~\ref{thm:fo2} we see that $\overline{K}$ is
$\FO^2$-definable.
\qed

%%%%%%%%%%%%%%%%%%%%%%%%%%%%%%%%%%%%%%%%%%%%%%%%%%%%%%%%%%%%%%%%%%%%%%

\subsection[The fragment $\FO^2 \cap \Pi_2$]{The fragment
  $\mathbf{\BFO^2 \cap \Pi_2}$}

Next, we discuss properties of closed unambiguous polynomials and
closed unambiguous monomials.

\begin{theorem}\label{thm:fo2pi2}
  Let $L \subseteq \Gamma^{\infty}$ be a regular language. The
  following assertions are equivalent:
  \begin{enumerate}
  \item\label{aaa:fo2pi2} $L$ is both $\FO^2$-definable and
    $\Pi_2$-definable.
  \item\label{bbb:fo2pi2} $L$ is $\FO^2$-definable and closed in
    the alphabetic topology.
  \item\label{ccc:fo2pi2} $L$ is the closure  in
    the alphabetic topology of some
    $\FO^2$-definable language.
  \end{enumerate}
\end{theorem}

\proof
The equivalence is the dual statement of the equivalence of
``\ref{aaa:fo2si2}'', ``\ref{bbb:fo2si2}'', and ``\ref{ddd:fo2si2}''
in Theorem~\ref{thm:fo2si2}.
\qed

Theorem~\ref{thm:fo2pi2} is not fully satisfactory since we do not
have any direct characterization in terms of polynomials.  We might
imagine that if $L$ is closed (and $L \in \FO^2 \cap \Pi_2$), then it
is a finite union of languages $K \cap B^{\im}$ where each $K \cap
B^{\im}$ is closed.  But this is not true:
Let $L = \Gamma^* a \cup \Gamma^{\omega}$, then $L$ is closed and in
$\FO^2 \cap \Pi_2$, but cannot be written in this form because $L =
\Gamma^* a$ is not closed.
We also note that the closure of a language $L \in \FO^2 \cap
\Sigma_2$ is not necessarily in $\Delta_2$. A counter-example is the
language $L = \Gamma^* abc$. By Lemma~\ref{lem:polclosure}, the
closure of $L$ is $\overline{L} = L \cup \Gamma^{\im}$ which is not
$\Sigma_2$-definable.

We have however a characterization when certain unambiguous monomials 
are closed: 

\begin{proposition}\label{p:clmono}
  Let $A_1^* a_1 \cdots A_k^* a_{k} A^\infty$ be unambiguous
   with $A_i \subseteq \oneset{a_i, \ldots, a_k}$ for all $1
  \leq i \leq k$ and let $P = A_1^* a_1 \cdots A_k^* a_{k} A^\infty
  \cap B^{\im}$ for some $B \subseteq A$.  The following assertions
  are equivalent:
  \begin{enumerate}
  \item\label{aaa:clmon} There is no $1 \leq i \leq k$ such
    that $B \subseteq \oneset{a_i, \ldots, a_k}\subseteq A_i$.
  \item\label{bbb:clmon} The unambiguous monomial $P= A_1^* a_1 \cdots
    A_k^* a_{k} A^\infty\cap B^{\im}$ is closed in the alphabetic
    topology.
  \end{enumerate}
\end{proposition}

\proof 
``\ref{aaa:clmon}~$\Rightarrow$~\ref{bbb:clmon}'': Assume by
contradiction that $P$ is not closed.  Let $\alpha \notin P$ with
$\im(\alpha) = C$ such that $\alpha $ is in the closure of $P$.  Then,
by Lemma~\ref{lem:polclosure}, there is some $1 \leq i \leq k$ such
that $\oneset{a_i, \ldots, a_k} \cup B \subseteq C \subseteq A_i$.
Thus, $\oneset{a_i, \ldots, a_k}= C = A_i$ since by hypotheses $A_i
\subseteq \oneset{a_i, \ldots, a_k}$.  Since $\alpha $ is in the
closure of $P$ we have $B \subseteq C= \oneset{a_i, \ldots, a_k}=
A_i$.  This is a contradiction to ``\ref{aaa:clmon}''.

``\ref{bbb:clmon}~$\Rightarrow$~\ref{aaa:clmon}'':
Assume by contradiction that $ B \subseteq \oneset{a_i, \ldots,
  a_k}\subseteq A_i$ for some $1 \leq i \leq k$. We have $a_1 \cdots
a_{i-1}(a_i \cdots a_k)^m B^\infty \cap B^{\im,} \subseteq P$ for all
$m\geq 1$ because $B \subseteq A$. As $P$ is closed and $ B \subseteq
\oneset{a_i, \ldots, a_k}$ we see $a_1 \cdots a_{i-1}(a_i \cdots
a_k)^{\omega} \in P$ and hence $ \oneset{a_i, \ldots, a_k}\subseteq
A$.  But this is a contradiction to the fact that $P$ is unambiguous
since $ \oneset{a_i, \ldots, a_k}\subseteq A_i \cap A$ implies that
$a_1 \cdots a_{i-1}(a_i \cdots a_k)^2$ has two different
factorizations.
\qed

\subsection[The relation between $\FO^{2}$ and $\Sigma_2 \cap
\Pi_2$]{The relation between $\mathbf{\BFO^{2}}$ and $\mathbf{\Sigma_2
    \cap \Pi_2}$}

For finite words we have the well-known theorem that
$\FO^2$-definability is equivalent to
$\Delta_2$-definability. However, this does not transfer to infinite
words, where $\Delta_2$ forms a proper subclass of $\FO^2$. Consider
$L= \oneset{a,b}^{\im}$, then $L$ is neither open nor closed, in
general. Hence $L \in \FO^2 \setminus (\Sigma_2\cup \Pi_2)$. The
result for finite words is therefore somewhat misleading.  The correct
translation for the general case is given in the following theorem,
which covers the situation for finite words by choosing $A =
\emptyset$.

\begin{theorem}\label{thm:lohengrin}
  For all $A \subseteq \Gamma$ the following assertions are
  equivalent:
  \begin{enumerate}
  \item\label{aaa:lohengrin} $L \cap A^{\im} $ is $\FO^2$-definable.
  \item\label{bbb:lohengrin} There are languages $L_\sigma \in \FO^2
    \cap \Sigma_2$ and $L_\pi \in \FO^2 \cap \Pi_2$ such that
    \begin{equation*}
      L \cap A^{\im} = L_\sigma \cap A^{\im} =  L_\pi \cap A^{\im}.
    \end{equation*}
  \item\label{ccc:lohengrin} There are languages $L_\sigma \in
    \Sigma_2$ and $L_\pi \in \Pi_2$ such that
    \begin{equation*}
      L \cap A^{\im} = L_\sigma \cap A^{\im}=  L_\pi \cap A^{\im}.
    \end{equation*}
  \end{enumerate}
\end{theorem}

\proof
``\ref{aaa:lohengrin}~$\Rightarrow$~\ref{bbb:lohengrin}'': By
Theorem~\ref{thm:fo2} we see that $L \cap A^{\im}$ is a finite union
of unambiguous monomials $A_1^* a_1 \cdots A_k^* a_k A^{\infty} \cap
A^{\im}$. We let $L_{\sigma}$ be the finite union of the monomials
$A_1^* a_1 \cdots A_k^* a_k A^{\infty}$; by
Theorem~\ref{thm:fo2si2} we obtain $L_{\sigma} \in \FO^2 \cap
\Sigma_2$.  Let $K$ be the complement of $L\cap A^{\im} $. Then $K$
and $K\cap A^{\im} $ are $\FO^2$-definable. Thus, $K\cap A^{\im} =
K_\sigma \cap A^{\im}$ for some $K_\sigma \in \FO^2 \cap
\Sigma_2$. Let $L_\pi$ be the complement of $K_\sigma$. Then $L_\pi
\in \FO^2 \cap \Pi_2$ and $L\cap A^{\im} = L_\pi \cap A^{\im}.$
``\ref{bbb:lohengrin}~$\Rightarrow$~\ref{ccc:lohengrin}'': Trivial.
``\ref{ccc:lohengrin}~$\Rightarrow$~\ref{aaa:lohengrin}'': If $L =
L_\sigma \cap A^{\im}$, then a slight modification of the proof for
Lemma~\ref{lem:Pol:LocHigh} shows that all idempotents in $\Synt(L)$
are locally top. Identically, if $L= L_\pi \cap A^{\im}$, then all
idempotents in $\Synt(L)$ are locally bottom. Thus $\Synt(L) \in \DA$,
and by Theorem~\ref{thm:fo2} we see that $L$ is $\FO^2$-definable.
\qed

%%%%%%%%%%%%%%%%%%%%%%%%%%%%%%%%%%%%%%%%%%%%%%%%%%%%%

\section{The fragment $\mathbf{\Delta_2 = \Sigma_2 \cap \Pi_2}$}\label{sec:delta2}

The first-order fragment $\Delta_2$ is the intersection of $\Sigma_2$
and $\Pi_2$. It is the largest subclass of $\Sigma_2$ (and also of
$\Pi_2$) which is closed under negation. Since over finite and
infinite words we have $\Sigma^\omega \not\in \Sigma_2$ and $\Sigma^*
\not\in \Pi_2$, we obtain different intersections $\Sigma_2 \cap
\Pi_2$ depending on whether we consider finite words, infinite words,
or simultaneously finite and infinite words.  In this section, we will
give characterizations of $\Delta_2$ for infinite words
$\Gamma^\omega$ and for finite and infinite words $\Gamma^\infty$. In
both settings, it will turn out that $\Delta_2$ is a strict subclass
of $\FO^2$.

\subsection{Clopen unambiguous monomials}

Languages in $\Sigma_2$ are open and languages in $\Pi_2$ are
closed. Hence, a language in $\Delta_2$ must be clopen in the
alphabetic topology. The first step towards a convenient
characterization of $\Delta_2$ is therefore a description of clopen
unambiguous monomials.

\begin{lemma}\label{lem:ClosedMonomial}
  Let $P = A_1^* a_1 \cdots A_k^* a_{k} A^\infty$ be an unambiguous
  monomial.  The following assertions are equivalent:
  \begin{enumerate}
  \item\label{aaa:ClosedMonomial} There is no $1 \leq i \leq k$ such
    that $ \oneset{a_i, \ldots, a_k}\subseteq A_i$.
  \item\label{bbb:ClosedMonomial} $P$ is closed in the alphabetic
    topology.
  \item\label{ccc:ClosedMonomial} $P$ is clopen in the alphabetic
    topology.
  \end{enumerate}
\end{lemma}

\proof
``\ref{aaa:ClosedMonomial}~$\Rightarrow$~\ref{bbb:ClosedMonomial}'':
By Lemma~\ref{lem:polclosure} (setting $A_{k+1} = A$) we see that the
closure of $P$ is:
\begin{equation*}
  \bigcup_{\oneset{a_i, \ldots, a_k}\subseteq B \subseteq A_i} 
  A_1^* a_1  \cdots A_{i-1}^* a_{i-1} A_i^\infty\cap B^{\im}.
\end{equation*}
Since there is no $ \oneset{a_i, \ldots, a_k}\subseteq A_i$ for $1
\leq i \leq k$, we see that this union is just $P$ itself. Therefore,
$P$ is closed.
``\ref{bbb:ClosedMonomial}~$\Rightarrow$~\ref{ccc:ClosedMonomial}'':
is clear, because $P$ is open.
``\ref{ccc:ClosedMonomial}~$\Rightarrow$~\ref{aaa:ClosedMonomial}'':
Assume by contradiction that $ \oneset{a_i, \ldots, a_k}\subseteq A_i$
for some $1 \leq i \leq k$. We have $a_1 \cdots a_{i-1}(a_i \cdots
a_k)^m \in P$ for all $m\geq 1$. As $P$ is closed we see $a_1 \cdots
a_{i-1}(a_i \cdots a_k)^{\omega} \in P$ and hence $ \oneset{a_i,
  \ldots, a_k}\subseteq A$.  But this is a contradiction to the fact
that $P$ is unambiguous since $ \oneset{a_i, \ldots, a_k}\subseteq A_i
\cap A$ implies that $a_1 \cdots a_{i-1}(a_i \cdots a_k)^2 \in P$ has
two different factorizations.
\qed

\begin{lemma}\label{lem:clopenLang}
  Let $L \subseteq \Gamma^{\infty}$ be a closed polynomial. For every
  unambiguous monomial $$P = A_1^* a_1 \cdots A_k^* a_k
  A^{\infty} \subseteq L$$ there exist closed unambiguous
  monomials $Q_1, \ldots, Q_{\ell}$ such that $P \subseteq Q_1 \cup
  \cdots \cup Q_{\ell} \subseteq L$, i.e., there exists a finite
  covering of $P$ with closed unambiguous monomials in $L$.
\end{lemma}

\proof
We start with a normalization procedure in which we begin with making
the last appearances of the letters in $A_i^*$ explicit.  We have $B^*
= \left( B \setminus \smallset{b} \right)^* \cup B^* b \left( B
  \setminus \smallset{b} \right)^*$ for every $b \in B$. This yields
the substitution rule of replacing $A_i^*$ in $P$ by $(A_i \setminus
\smallset{a})^*$ and also by $A_i^* a (A_i \setminus \smallset{a})^*$
which gives two new monomials.  After iterating this substitution rule
a finite number of times, we obtain unambiguous monomials of the form
$P'_i = B_1^* b_1 \cdots B_s^* b_s A^{\infty}$ such that $P = \bigcup
P'_i$ and $B_i \subseteq \oneset{b_i, \ldots, b_s}$ for every $1 \leq
i \leq s$. In the next phase of the normalization procedure we make
the first appearances of the letters in $A^{\infty}$ explicit. We have
$B^\infty = \left( B \setminus \smallset{b} \right)^\infty \cup \left(
  B \setminus \smallset{b} \right)^* b B^\infty$ for every $b \in
B$. As above, this yields a substitution rule and after a finite
number of applications to the $P'_i$ we obtain unambiguous monomials
of the form $P''_i = B_1^* b_1 \cdots B_s^* b_s B_{s+1}^* b_{s+1}
\cdots B_t^* b_t A^{\infty}$ such that $P = \bigcup P''_i$ and the
following properties hold:
\begin{itemize}
\item $B_i \subseteq \oneset{b_i, \ldots, b_t}$ for every $1 \leq i
  \leq s$.
\item $\oneset{b_i, \ldots, b_t} \not\subseteq B_i$ for all $s+1 \leq
  i \leq t$.
\item $A = \oneset{b_{s+1}, \ldots, b_t}$.
\end{itemize}
It suffices to prove the lemma for $P = B_1^* b_1 \cdots B_s^* b_s
B_{s+1}^* b_{s+1} \cdots B_t^* b_t A^{\infty}$ with the above
properties.
If $P$ is not closed, then by Lemma~\ref{lem:ClosedMonomial} there
exists $1 \leq i \leq s$ such that $B_i \supseteq \oneset{b_i, \ldots,
  b_t}$, and hence $A \subseteq B_i = \oneset{b_i, \ldots, b_t}$ due
to the normalization procedure. We fix the minimal index $i$ with this
property. 

Next, we use a Ramsey argument. Let $L$ be strongly recognized by $h :
\Gamma^* \to M$ and let $r=r(M)$ be the Ramsey number such that every
complete edge-colored graph with $r$ nodes and using at most $\abs{M}$
colors contains a monochromatic triangle. We have $B_i^* =
(B_i\setminus\smallset{b_j})^* \cup (B_i\setminus\smallset{b_j})^* b_j
B_i^*$ and $B_i\setminus\smallset{b_j}$ is no longer a superset of
$\oneset{b_i, \ldots, b_t}$. Therefore, we only have to
consider the case where we replace the factor $b_{i-1} B_i^* b_i$ in
$P$ by $b_{i-1} (B_i \setminus \smallset{b_j})^* b_j B_i^* b_i$ for
some $i \leq j \leq t$. Repeating this procedure we are left with a
situation where we have replaced $ b_{i-1} B_i^* b_i$ in $P$ by
$b_{i-1} R^r B_i^* b_i$ in $P$ where
\begin{equation*}
  R = (B_i \setminus \smallset{b_i})^* b_i 
  (B_i \setminus \smallset{b_{i+1}})^* b_{i+1}
  \cdots (B_i \setminus\smallset{b_t})^* b_t.
\end{equation*}
Note that the resulting monomial $\tilde{P}$ is unambiguous and that the
alphabet of every word in $R$ is $B_i = \oneset{b_i, \ldots, b_t}$.

Now consider $\alpha = u v_1 \cdots v_r \in B_1^* b_1 \cdots B_{i-1}^*
b_{i-1} R^r$, with $v_j \in R$ for all $1\leq j \leq r$.  By the
choice of $r$ being the Ramsey number for triangles we find some $j_1
\leq j_2 < j_3$ such that $h(v_{j_1} \cdots v_{j_2}) = h(v_{j_2 + 1}
\cdots v_{j_3}) = h(v_{j_1} \cdots v_{j_3})$ is idempotent in the
monoid $M$.  Since $L$ is closed we see that
\begin{equation*}
  uv_1 \cdots v_{j_1 - 1} (v_{j_1} \cdots v_{j_2}) ^{\omega} \in L.
\end{equation*}
Indeed, for each prefix $w_m = u v_1 \cdots v_{j_1 - 1}
(v_{j_1} \cdots v_{j_2})^m$ we have $\alp(v_{j_1}) = \oneset{b_i,
  \ldots, b_t} = B_i$ and $w_m b_i \cdots b_{t} \in P \subseteq L$.
 
Since $L$ is open, there is some $m$ such that $w_m B_i^\infty
\subseteq L$. This follows again because $\alp(v_{j_1}) = B_i$. Since
$h$ strongly recognizes $L$ and since $h(w_m) = h(uv_1 \cdots
v_{j_2})$ by idempotency of $h(v_{j_1} \cdots v_{j_2})$, we have $uv_1
\cdots v_{j_2} B_i^\infty \subseteq L$. In particular, $u v_1 \cdots
v_r B_i^\infty \subseteq L$. 

This is true for all $\alpha \in B_1^*
b_1 \cdots B_{i-1}^* b_{i-1} R^r$, hence
\begin{equation*}
  B_1^* b_1  \cdots B_{i-1}^* b_{i-1} R^r B_i^\infty \subseteq L.
\end{equation*}
By construction, $Q = B_1^* b_1 \cdots B_{i-1}^* b_{i-1} R^r B_i^\infty$ is a
closed unambiguous monomial and due to the normalization, we have
$B_i^* b_i \cdots B_t^* b_t A^{\infty} \subseteq B_i^{\infty}$ and hence
$P \subseteq Q$.
\qed

\subsection{Arrow languages and deterministic languages}

The results of this section are very similar to results on
deterministic and \codet languages which can be found in~\cite{pp04},
too. Moreover, the conditions in Proposition~\ref{prop:arl} and
Proposition~\ref{p:detdis} can be complemented by several other
equivalent characterizations, see
e.g.~\cite[Theorem~VI.3.7]{pp04}. One of them is the class of finite
Boolean combinations of regular Cantor-open languages and another one
is in terms of the second level of the Borel hierarchy over the Cantor
topology.

We write $s \gRop t$ for monoid elements $s,t \in M$ if there exist
$x,y \in M$ such that $s = ty$ and $t = sx$, i.e., if the
\emph{right-ideals} $sM$ and $tM$ are equal.  The relation $\gR$ is
one of Green's relations, see e.g.~\cite{pin86}.

\begin{lemma}\label{l:dRele}
  Let $L \subseteq \Gamma^{\infty}$ be a deterministic language which
  is strongly recognized by some surjective homomorphism $h : \Gamma^*
  \to M$ onto a finite monoid $M$.  Let $s,e,t,f,x,y, \in M$ such that
  $(s,e), (t,f)$ are linked pairs and $s = ty$ and $t = sx$
  (thus, $s \gRop t$). Assume
  that
  $$[s][e]^{\omega} \cap L \cap \Gamma^{\omega} \neq \emptyset.$$
  Then we have $[t][yexf]^{\omega} \subseteq L$.
\end{lemma}

\proof
Let $s_0,e_0,f_0,x_0,y_0 \in \GA^*$ be words which are mapped to the
corresponding elements in $s,e,f,x,y \in M$.  We choose $e_0 \neq 1$
nonempty, which we can do due to the assumption. Since $L$ is
deterministic, there exists a set $W \subseteq \Gamma^*$ such that $L
\cap \GA^\oo = \overrightarrow{W} \cap \GA^\oo$.  We are going to
construct sequences of words $s_n \in [s]\big([xf][ye]\big)^n$ and
$w_n \in W$ for $n \in \N$ such that
$$s_0 < w_0 < s_1 < w_1 < s_2 < w_2 < \cdots$$
where $<$ denotes the strict prefix order on words.  Thus, the limit
defines an infinite word $\alpha$ such that $\alpha \in
[s]\big([xf][ye]\big)^{\omega} \cap \overrightarrow{W}$.  In
particular, $\alpha \in L$. Moreover, since $sxf = t$ we have $\alpha
\in [t][yexf]^\omega \cap L$ and hence $[t][yexf]^\omega \subseteq L$
due to strong recognition.

Thus, it is enough to define the sequences $s_n$ and $w_n$ for $n \in
\N$ as above. The condition $s_0 \in [s]\big([xf][ye]\big)^0$ is
satisfied by definition.  Let $n\in \N$.  Inductively, we may assume
that $w_k$ and $s_m$ are defined as desired for $k < n$ and $m \leq
n$.  We are going to define $w_n$ and $s_{n+1}$.  Infinitely many
prefixes of $s_n x_0 f_0 y_0 e_0^\oo$ are in $W$, because $s_n x_0 f_0
y_0 e_0^\oo \in [s][e]^\oo \subseteq L$. Thus we find $w_n \in W$ and
$\ell \geq 1$ such that
$$s_n < w_n < s_{n+1} = s_n x_0 f_0 y_0 e_0^\ell.$$
By induction we see that $s_{n+1} \in [s]\big([x f][y e]\big)^{n+1}$
because $x_0 f_0 y_0 e_0^\ell \in [x f][y e]$ since $e^2=e$.
\qed

\begin{proposition}\label{prop:arl}
  Let $L \subseteq \Gamma^\infty$ be strongly recognized by some
  surjective homomorphism $h : \Gamma^* \to M$ onto a finite monoid
  $M$. Define
  \begin{equation*}
    W = \bigcup \set{[s] \subseteq \Gamma^*}{[s][e]^\omega \subseteq L 
      \text{ for some linked pair } (s,e)}.
  \end{equation*}
  Then the following four assertions are equivalent:
  \begin{enumerate}
  \item\label{aaa:arl} $L = \overrightarrow{W}$.
  \item\label{ccc:arl} For all linked pairs $(s,e)$,
    $(t,f)$ with $s \gR t$ we have
    \begin{equation*}
      [s][e]^\omega \subseteq L \ \Leftrightarrow \ [t][f]^\omega \subseteq L.
    \end{equation*}
  \item\label{ddd:arl} For every linked pair $(s,e)$ we
    have
    \begin{equation*}
      [s][e]^\omega \subseteq L \ \Leftrightarrow \ [s] \subseteq L.
    \end{equation*}
  \item\label{bbb:arl} Both $L$ and its complement are
    arrow languages.
  \end{enumerate}
\end{proposition}

\proof
``\ref{aaa:arl}~$\Rightarrow$~\ref{ccc:arl}'': Let $[s] \subseteq W$
and let $(t,f)$ be a linked pair with $s \gR t$.  It is enough to show
$[t][f]^\omega \subseteq L$. If $s = t$, then $[t][f]^\omega \subseteq
\overrightarrow{W} = L$. For $s \neq t$ we find $x \neq 1 \neq y$ with
$s = ty$ and $t = sx$. It follows that $[s][xy]^\omega \cap L \cap
\Gamma^\omega \neq \emptyset$.  Lemma~\ref{l:dRele} yields
$[t][yexf]^\omega \subseteq L$ for $e = xy$. But then $[t] \subseteq
W$ and $[t][f]^\omega \subseteq \overrightarrow{W} = L$.

``\ref{ccc:arl}~$\Rightarrow$~\ref{ddd:arl}'': If $[s][e]^\omega
\subseteq L$ then by ``\ref{ccc:arl}'' we have $[s][1]^\omega
\subseteq L$. Since $[s] \subseteq [s][1]^\omega$, it follows $[s]
\subseteq L$. Conversely, if $[s] \subseteq L$, then strong
recognition yields $[s][1]^\omega \subseteq L$; and hence
$[s][e]^\omega \subseteq L$ by ``\ref{ccc:arl}''.

``\ref{ddd:arl}~$\Rightarrow$~\ref{bbb:arl}'': The condition is
symmetric in $L$ and its complement. Therefore it is enough to show
that $L$ is an \arr language. We show $L = \overrightarrow{L \cap
  \Gamma^*}$. Let $[s][e]^\omega \subseteq L$. Then, by
``\ref{ddd:arl}'', we see that $[s] \subseteq L$ and hence
$[s][e]^\omega \subseteq \overrightarrow{[s]} \subseteq
\overrightarrow{L \cap \Gamma^*}$. For the other inclusion, let
$\alpha \in \overrightarrow{L \cap \Gamma^*}$. Then $\alpha \in
\overrightarrow{[s]}$ for some $s \in M$ with $[s] \cap L \neq
\emptyset$. We can find a linked pair $(s,e)$ such that $\alpha \in
[s][e]^\omega$.  By strong recognition, $[s] \subseteq [s][1]^\omega
\subseteq L$. By ``\ref{ddd:arl}'' we conclude $[s][e]^\omega
\subseteq L$ and $\alpha \in L$.

``\ref{bbb:arl}~$\Rightarrow$~\ref{aaa:arl}'': Since $L$ is an arrow
language, it is enough to show $L \cap \GA^* = W$. The inclusion
$L \cap \GA^* \subseteq W$ is trivial.  For the converse assume by
contradiction $[s]\cap L = \emptyset$, but $[s][e]^\oo \subseteq L $
for some linked pair $(s,e)$. Then $[s]\subseteq \Gamma^* \setminus
L$. Since the complement of $L$ is an arrow language, we have
$[s][e]^\oo \subseteq \overrightarrow{[s]} \subseteq
\overrightarrow{\Gamma^* \setminus L} = \Gamma^\infty \setminus L$,
which is a contradiction to $[s][e]^\oo \subseteq L $. Thus, $W
\subseteq L \cap \GA^*$.
\qed

The following result yields a simple proof for a \emph{Landweber type}
result in the special case of deterministic and \codet languages.

\begin{proposition}\label{p:detdis}
  Let $L \subseteq \Gamma^{\omega}$ be a deterministic language which
  is strongly recognized by some surjective homomorphism $h : \Gamma^*
  \to M$ onto a finite monoid $M$. Let
  \begin{equation*}
    W = \bigcup \set{[s] \subseteq \Gamma^*}{[s][e]^\omega \subseteq L 
      \text{ for some linked pair } (s,e)}
  \end{equation*}
  and $U = \Gamma^* \setminus W$. Then $\overrightarrow{W} \cup
  \overrightarrow{U} = \Gamma^\infty$ and $\overrightarrow{W} \cap
  \overrightarrow{U} = \emptyset$, i.e., $\Gamma^\infty$ is a disjoint
  union of $\overrightarrow{W}$ and $\overrightarrow{U}$. Moreover,
  $\overrightarrow{W} \cap \Gamma^\omega = L$ if and only if $L$
  is complement-deterministic, too. 
\end{proposition}

\proof
Clearly, $\overrightarrow{W} \cup \overrightarrow{U} =
\Gamma^\infty$. Assume by contradiction that there is some $\alpha \in
\overrightarrow{W} \cap \overrightarrow{U}$. Then $\alpha \in
\Gamma^\omega$ with $\alpha \in \overrightarrow{[s]} \cap
\overrightarrow{[t]}$ such that $[s] \subseteq W$ and $[t] \subseteq
U$. Using the usual application of Ramsey's Theorem at those prefixes
belonging to $[s]$ or $[t]$, respectively, we see that for some linked
pairs $(s,e)$, $(t,f)$ we have $\alpha \in [s][e]^\omega$ and $\alpha
\in [t][f]^\omega$. We have $s = ty$ and $t = sx$ with $x \neq 1 \neq
y$ because $s \neq t$ as $U \cap W = \emptyset$. Since $[s][e]^\omega
\cap L \cap \Gamma^\omega \neq \emptyset$, by
Lemma~\ref{l:dRele} we have $[t][yexf]^\omega \subseteq
L$. This contradicts $[t] \subseteq U = \Gamma^* \setminus W$.

For the second statement of the proposition: If $\overrightarrow{W}
\cap \Gamma^\omega = L$, then by the first statement of this
proposition we have $\Gamma^\omega \setminus L = \overrightarrow{U}
\cap \Gamma^\omega$, i.e., $L$ is complement-deterministic.

For the converse, let $L$ be complement-deterministic. Clearly, $L
\subseteq \overrightarrow{W} \cap \Gamma^\omega$. Assume by
contradiction that there is some $\alpha \in \overrightarrow{W}
\setminus L$ for some $\alpha \in \Gamma^\omega$.  Then $\alpha \in
\overrightarrow{[s]} \cap [t][f]^\omega$ for $[s] \subseteq W$ and
$(t,f)$ is a linked pair with $[t][f]^\omega \subseteq \Gamma^\omega
\setminus L$. We have $s = ty$ and $t = sx$ for some $x,y \in M$. By
definition of $W$, we find a linked pair $(s,e)$ such that
$[s][e]^\omega \subseteq L$.  We have $[s][e]^\omega \cap L \cap
\Gamma^\omega \neq \emptyset$ and $[t][f]^\omega \cap (\Gamma^\infty
\setminus L) \cap \Gamma^\omega \neq \emptyset$. Since both $L$ and
$\Gamma^\infty \setminus L$ are deterministic, we can apply
Lemma~\ref{l:dRele} and obtain $[t][yexf]^\omega \subseteq
L$ and $[s][xfye]^\omega \subseteq \Gamma^\omega \setminus L$.  This
is a contradiction to strong recognizability, since $[t][yexf]^\omega
\cap [s][xfye]^\omega \neq \emptyset$.
\qed

\subsection[Various characterizations of $\Delta_2$]{Various
  characterizations of $\mathbf{\Delta_2}$}

We are now ready to characterize $\Delta_2$-definable languages $L
\subseteq \Gamma^\infty$ over finite and infinite words.

\begin{theorem}\label{thm:delta2}
  Let $L \subseteq \Gamma^{\infty}$ be a regular language. The
  following assertions are equivalent.
  \begin{enumerate}
  \item\label{aaa:delta2} $L$ is $\Delta_2$-definable.
  \item\label{bbb:delta2} $L$ is $\FO^2$-definable and $L$ is
    clopen in the alphabetic topology.
  \item\label{aba:delta2} $L$ is a finite union of unambiguous closed
    monomials $A_1^* a_1 \cdots A_k^* a_{k} A^\infty$, i.e., there is
    no $1 \leq i \leq k$ such that $ \oneset{a_i, \ldots,
      a_k}\subseteq A_i$.
  \item\label{ccc:delta2} $\Synt(L) \in \DA$ and for
    all linked pairs $(s,e)$, $(t,f)$ with $s \gRop t$ (i.e., there
    exist $x,y \in \Synt(L)$ such that $s = ty$ and $t = sx$) we have
    \begin{equation*}
      [s][e]^{\omega} \subseteq L \ \Leftrightarrow \ 
      [t][f]^{\omega} \subseteq L.
    \end{equation*}
  \item\label{ddd:delta2} $L$ is weakly recognized by $h : \Gamma^*
    \to M$ for some $M \in \DA$, and for all linked pairs $(s,e)$,
    $(t,f)$ with $s \gRop t$ in $M$ we have $[s][e]^{\omega} \subseteq
    L$ if and only if $[t][f]^{\omega} \subseteq L$.
  \item\label{eee:delta2} $\Synt(L) \in \DA$ and both $L$ and its
    complement $\Gamma^{\infty} \setminus L$ are arrow languages.
  \end{enumerate}
\end{theorem}

\proof
``\ref{aaa:delta2}~$\Rightarrow$~\ref{bbb:delta2}'': By
Theorem~\ref{thm:sigma2} and its dual version for $\Pi_2$, we see that
$\Synt(L) \in \DA$ and that $L$ is clopen in the alphabetic topology.
From Theorem~\ref{thm:fo2} it follows that $L$ is $\FO^2$-definable.

``\ref{bbb:delta2}~$\Rightarrow$~\ref{aba:delta2}'': By
Theorem~\ref{thm:fo2si2}, $L$ is a finite union of unambiguous
monomials. Property ``\ref{aba:delta2}'' now follows by
Lemma~\ref{lem:clopenLang} and Lemma~\ref{lem:ClosedMonomial}.

``\ref{aba:delta2}~$\Rightarrow$~\ref{aaa:delta2}'':
Theorem~\ref{thm:fo2si2} and Theorem~\ref{thm:fo2pi2}.

``\ref{bbb:delta2}~$\Rightarrow$~\ref{ccc:delta2}'': By
Theorem~\ref{thm:fo2}, we see that $\Synt(L) \in \DA$. Suppose
$[s][e]^{\omega} \subseteq L$ and let $s = ty$ and $t = sx$. Since $L$
is closed we see that $[s][exfy]^{\omega} \subseteq L$ and by strong
recognition we conclude $[t][fyex]^{\omega} \subseteq L$. Let $A =
\bigcup \set{\alp(v)}{v \in [f]}$.  Since $L$ is open and by strong
recognition, there exists $r \in \N$ such that $[t][fyex]^r A^{\infty}
\subseteq L$. Moreover, $t = tfyex$ and thus, $[t] A^{\infty}
\subseteq L$. In particular, $[t][f]^{\omega} \subseteq L$ because
$[f] \subseteq A^*$.

``\ref{ccc:delta2}~$\Rightarrow$~\ref{ddd:delta2}'': Trivial with $M =
\Synt(L)$ and $h = h_L$.

``\ref{ddd:delta2}~$\Rightarrow$~\ref{bbb:delta2}'': If $\alpha \in
[s][e]^\omega \cap [t][f]^\omega$ for linked pairs $(s,e), (t,f)$,
then $s \gR t$. Hence $[s][e]^\omega \subseteq L$ and $[s][e]^\omega
\cap [t][f]^\omega \neq \emptyset$ implies $[t][f]^\omega \subseteq
L$. In particular, $h$ strongly recognizes $L$.

Definability in $\FO^2$ follows by Theorem~\ref{thm:fo2}. By symmetry,
it suffices to show that $L$ is open. Let $\alpha \in [s][e]^{\omega}
\subseteq L$ for some linked pair $(s,e)$ and write $\alpha = u \beta$
with $u \in [s]$ and $\beta \in [e]^{\omega} \cap \cpx{A}$ for some $A
\subseteq \Gamma$. Let $v \leq \beta$ be a prefix such that $v \in
[e]$ and $\alp(v) = \alp(\beta)$. We want to show $uv A^{\infty}
\subseteq L$. Consider $uv\gamma \in \Gamma^{\infty}$ where $\gamma
\in A^{\infty}$. We have $uv\gamma \in [t][f]^{\omega}$ for some
linked pair $(t,f)$.  Let $v' \leq \gamma$ such that $uvv' \in
[t]$. Since $\Synt(L) \in \DA$ we have $v v' v \in [e]$ and $s = t
\cdot h(v)$. Together with $t = s \cdot h(v')$ it follows $s \gRop t$
and by ``\ref{ccc:delta2}'' we obtain $uv\gamma \in [t][f]^{\omega}
\subseteq L$.

``\ref{ccc:delta2}~$\Leftrightarrow$~\ref{eee:delta2}'': 
This equivalence follows from Proposition~\ref{prop:arl}. 
\qed

\begin{corollary}\label{cor:clopen:det}
  Let $L \subseteq \Gamma^{\infty}$ be a regular language such that
  $\Synt(L) \in \DA$. The following assertions are equivalent:
  \begin{enumerate}
  \item\label{aaa:clopen:det} $L$ is clopen in the alphabetic topology.
  \item\label{bbb:clopen:det} Both $L$ and its complement
    $\Gamma^{\infty} \setminus L$ are arrow languages.
  \end{enumerate}
\end{corollary}

\proof
The statement follows from the equivalence of ``\ref{bbb:delta2}'' and
``\ref{eee:delta2}'' in Theorem~\ref{thm:delta2} since by
Theorem~\ref{thm:fo2} the language $L$ is $\FO^2$-definable if and
only if $\Synt(L) \in \DA$.
\qed

\begin{remark}
  The statement of Corollary~\ref{cor:clopen:det} does not need to
  hold outside the variety $\DA$. For example the aperiodic language
  $L = (ab)^\oo \cup (ab)^* a \subseteq \oneset{a,b}^\infty$ is an
  arrow language and its complement is also an arrow language, but it
  is not open.
\end{remark}

\subsection[The intersection of $\Sigma_2$ and $\Pi_2$ over infinite
words]{The intersection of $\mathbf{\Sigma_2}$ and $\mathbf{\Pi_2}$
  over infinite words}

The next corollary gives a characterization of the fragment $\Delta_2$
for $\omega$-languages, i.e., we consider the intersection of
$\Sigma_2$ and $\Pi_2$ over infinite words (instead of finite and
infinite words). Note that the language $\Gamma^\omega \subseteq
\Gamma^\infty$ of all infinite words is $\Pi_2$-definable, but not
$\Sigma_2$-definable as a subset of $\Gamma^\infty$.

\begin{corollary}\label{cor:d2o}
  Let $L \subseteq \Gamma^{\oo}$ be an $\omega$-regular language. The
  following assertions are equivalent:
  \begin{enumerate}
  \item\label{aaa:d2o} $L \in \Pi_2$ and there exists a language
    $L_\sigma \in \Sigma_2$ such that $L = L_\sigma \cap \GA^\oo$.
  \item\label{aab:d2o} There exist languages $L_\sigma \in
    \Sigma_2$ and $L_\pi \in \Pi_2$ such that $L = L_\sigma \cap
    \Gamma^\omega = L_\pi \cap \Gamma^\omega$.
  \item\label{bbb:d2o} $\Synt(L) \in \DA$ and $L$ is
    deterministic and complement-deterministic.
  \item\label{ccc:d2o} There exists a language $L_{\delta} \in
    \Delta_2$ such that $L = L_{\delta} \cap \Gamma^{\omega}$.
  \end{enumerate}
\end{corollary}

\proof
``\ref{aaa:d2o}~$\Leftrightarrow$~\ref{aab:d2o}'':
Trivial, since $L = L_\pi \cap \Gamma^\omega$ is $\Pi_2$-definable.

``\ref{aab:d2o}~$\Rightarrow$~\ref{bbb:d2o}'': By
Theorem~\ref{thm:lohengrin} we see that $L$ is $\FO^2$-definable and
by Theorem~\ref{thm:fo2} we conclude $\Synt(L) \in \DA$.  The
complement of $L_\pi$ is $\Sigma_2$-definable, hence $L_\pi$ is closed
by Theorem~\ref{thm:sigma2}. Therefore, $L= L_\pi \cap \Gamma^\omega$
is closed too. By Corollary~\ref{cor:closedisdet} it follows that $L$
is deterministic.  Symmetrically, we deduce that $\Gamma^{\omega}
\setminus L$ is also deterministic.

``\ref{bbb:d2o}~$\Rightarrow$~\ref{ccc:d2o}'': Let $W
= \bigcup \set{[s] \subseteq \Gamma^*}{[s][e]^\omega \subseteq L
  \text{ for some linked pair } (s,e)}$ and set $L_\delta =
\overrightarrow{W}$. By Proposition~\ref{p:detdis} we have $L
= L_\delta \cap \Gamma^\omega$. Moreover, both $L_\delta$ and its
complement are arrow languages. Since $\Synt(L_\delta) = \Synt(L)$ we
can apply Theorem~\ref{thm:delta2} and conclude $L_\delta \in
\Delta_2$.

``\ref{ccc:d2o}~$\Rightarrow$~\ref{aab:d2o}'': Trivial
with $L_\sigma = L_\pi = L_\delta$.
\qed

\subsection{On the construction of examples}

In this section, we relate different classes of linked pairs with the
languages recognized by these classes. The different classes come from
several acceptance conditions of homomorphisms onto finite monoids
such as weak or strong recognition.  For monoids in $\DA$, the results
in the previous sections allow us to deduce non-trivial properties of
the languages recognized by the respective classes of linked pairs.

Let $h : \Gamma^* \to M$ be a surjective homomorphism onto a finite
monoid $M$.  By definition of weak recognition, for every linked pair
$(s,e)$ the language $[s][e]^\omega$ is weakly recognized by $h$ and
every language which is weakly recognized by $h$ is a union of such
languages. We say that two linked pairs $(s,e)$, $(t,f)$ are
\emph{conjugated}, if $e = xy$, $f = yx$, and $t = sx$ for some $x,y
\in M$. It is easy to verify that conjugacy forms an equivalence
relation on the set of linked pairs and that $[s][e]^\omega \cap
[t][f]^\omega \neq \emptyset$ if and only if the linked pairs $(s,e)$
and $(t,f)$ are conjugated. We define for a linked pair $(s,e)$ the
class $[s,e]$ as a language by:
\begin{equation*}
  [s,e] \;=\; \bigcup \set{[t][f]^\omega}{(s,e) \text{ and } 
    (t,f) \text{ are conjugated}} \;\subseteq\; \Gamma^\infty.
\end{equation*}
The language $[s,e]$ is strongly recognized by $h$; and every
language, which is strongly recognized by $h$, is a union of such
languages.

The set $\overrightarrow{[s]}$ is an arrow language which is
weakly recognized by $h$ since 
\begin{equation*}
  \overrightarrow{[s]} \;=\; \bigcup \set{[s][f]^\omega}{%
    (s,f) \text{ is a linked pair for some } f \in M}
\end{equation*}
If an arrow language $L \subseteq \Gamma^\infty$ is weakly recognized
by $h$, then $L$ is a union of languages of the form
$\overrightarrow{[s]}$ since $L = \overrightarrow{L \cap \Gamma^*}$
and $L \cap \Gamma^* = \bigcup \set{[s]}{[s] \cap L \neq \emptyset}$.
In general, $\overrightarrow{[s]}$ is not strongly recognized by $M$.

For every $s \in M$ we denote by $\gR_s$ the $\gR$-class of $s$, i.e.,
${\gR_s} = \set{t \in M}{sM = tM}$. We have
\begin{equation*}
  \overrightarrow{[\gR_s]} = \bigcup
  \set{[s,e]}{\text{there exists } e\in M \text{ such that } 
    (s,e) \text{ is a linked pair}}.
\end{equation*}
By Proposition~\ref{prop:arl}, both $\overrightarrow{[\gR_s]}$ and its
complement are arrow languages which are strongly recognized by $h$.
Conversely, if $L$ and its complement are arrow languages which are
strongly recognized by $h$, then $L$ is a union of languages of the
form $\overrightarrow{[\gR_s]}$. Moreover, as shown in the proof of
Theorem~\ref{thm:delta2}, if $L$ and its complement are arrow
languages and if $L$ is weakly recognized by $h$, then, in fact, $L$ is
strongly recognized by $h$.

Therefore, for some given $h : \Gamma^* \to M$ we find examples as
follows:
\begin{itemize}
\item $[s][e]^\omega$ are  languages which are
  weakly recognizable by $h$.
\item $[s,e]$ are languages which are strongly
  recognizable by $h$.
\item $\overrightarrow{[s]}$ are arrow languages which are weakly
  recognizable by $h$.
\item $\overrightarrow{[\gR_s]}$ are \arr
  languages whose complement is also an arrow language, and which are
  strongly recognizable by $h$.
\end{itemize}

More concretely: If $M \in \DA$ then, by Theorem~\ref{thm:fo2}, the
languages which are strongly recognizable by $h$ are
$\FO^2$-definable, but by Example~\ref{ex:nwds} weak recognition is
not enough to guarantee $\FO^2$-definability.  By
Theorem~\ref{thm:delta2}, languages $L$ are $\Delta_2$-definable if
they are strongly recognizable by $h$ and if both, $L$ and
$\Gamma^\infty \setminus L$ are arrow languages.

Therefore we can produce examples along the following line: We start
with some linked pair $(s,e)$, this yields $[s][e]^\omega$ which is
weakly recognizable and $[s,e]$ which is strongly recognized by $h$.
The \arr language $\overrightarrow{[s]}$ is incomparable with $[s,e]$,
in general.  By definition, $\overrightarrow{[s]}$ is a deterministic
language.  Moving to $\overrightarrow{[\gR_s]}$ yields an \arr
language, where its complement is an \arr language, too.  We have:
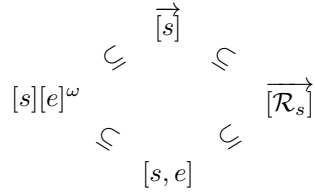
\begin{center}
  \begin{tikzpicture}
    \draw (0,0) node (se) {$[s][e]^\omega$};
    \draw (1.6,1) node (sarr) {$\overrightarrow{[s]}$};
    \draw (1.6,-1) node (seconj) {$[s,e]$};
    \draw (3.2,0) node (Rs) {$\overrightarrow{[\gR_s]}$};
    \path (se) -- (sarr) node[midway,sloped] {$\subseteq$};
    \path (sarr) -- (Rs) node[midway,sloped] {$\subseteq$};
    \path (se) -- (seconj) node[midway,sloped] {$\subseteq$};
    \path (seconj) -- (Rs) node[midway,sloped] {$\subseteq$};
  \end{tikzpicture}
\end{center}

\begin{example}\label{ex:big}
  Let $\Gamma = \oneset{a,b,c}$ and $P = c^* a \Gamma^* b \Gamma^* c$.
  The syntactic monoid of $P$ is in $\DA$, because $P$ is
  $\FO^2$-definable. We can write $\Synt(P) = \oneset{1,a,b,ab, c,
    ac,bc,abc}$ where the elements correspond to minimal length
  representatives of the classes induced by the syntactic
  congruence. To see this, observe that $P = c^* a \Gamma^* b \Gamma^*
  \cap \Gamma^* c$.  The syntactic monoid of $c^* a \Gamma^* b
  \Gamma^*$ has just the four elements in $\oneset{1,a,b,ab}$. For
  $\Synt(P)$ we copy these classes and add the information whether it
  represents a word ending in $c$.  All elements of $\Synt(P)$ are
  idempotent and its egg-box representation (see e.g.~\cite{pin86})
  is depicted in Figure~\ref{fig:exbig}.
  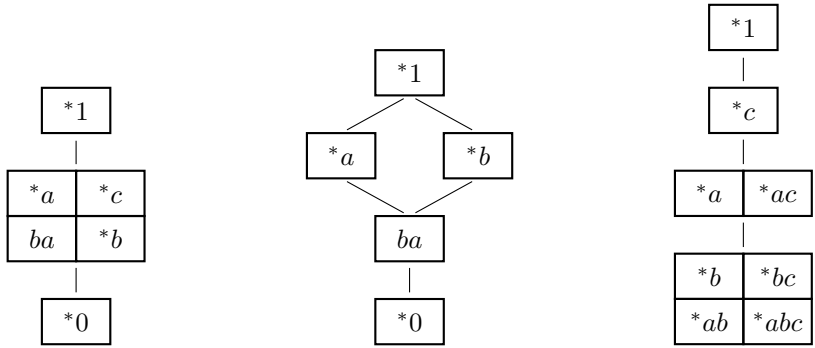
\begin{figure}[ht]
    \centering
    \subfigure[Monoid $M$ in Example~\ref{ex:damon}]{ \ \ \qquad 
    \begin{tikzpicture}
      \node(1) at (0,0) [thick,rectangle,draw,inner sep=0cm,minimum height=.6cm,minimum width=.9cm] {$^*1$};
      \node(a) at (-.45,-1.1) [thick,rectangle,draw,inner sep=0cm,minimum height=.6cm,minimum width=.9cm] {$^*a$};
      \node(c) at (.45,-1.1) [thick,rectangle,draw,inner sep=0cm,minimum height=.6cm,minimum width=.9cm] {$^*c$};
      \node(ba) at (-.45,-1.7) [thick,rectangle,draw,inner sep=0cm,minimum height=.6cm,minimum width=.9cm] {$ba$};
      \node(b) at (.45,-1.7) [thick,rectangle,draw,inner sep=0cm,minimum height=.6cm,minimum width=.9cm] {$^*b$};
      \node(0) at (0,-2.8) [thick,rectangle,draw,inner sep=0cm,minimum height=.6cm,minimum width=.9cm] {$^*0$};
      \draw[-,thin,shorten >=2.5pt,shorten <=2.5pt] (1) -- (0,-.8);
      \draw[-,thin,shorten >=2.5pt,shorten <=2.5pt] (0,-2.0) -- (0,-2.5);
    \end{tikzpicture} \ \ \qquad
    \label{fig:exdamonM}
    }~~\subfigure[Monoid $N$ in Example~\ref{ex:damon}]{ \qquad
    \begin{tikzpicture}
      \node(1) at (0,0) [thick,rectangle,draw,inner sep=0cm,minimum height=.6cm,minimum width=.9cm] {$^*1$};
      \node(a) at (-.9,-1.1) [thick,rectangle,draw,inner sep=0cm,minimum height=.6cm,minimum width=.9cm] {$^*a$};
      \node(b) at (.9,-1.1) [thick,rectangle,draw,inner sep=0cm,minimum height=.6cm,minimum width=.9cm] {$^*b$};
      \node(ba) at (0,-2.2) [thick,rectangle,draw,inner sep=0cm,minimum height=.6cm,minimum width=.9cm] {$ba$};
      \node(0) at (0,-3.3) [thick,rectangle,draw,inner sep=0cm,minimum height=.6cm,minimum width=.9cm] {$^*0$};
      \draw[-,thin,shorten >=2.5pt,shorten <=2.5pt] (0,-0.3) -- (-.9,-.8);
      \draw[-,thin,shorten >=2.5pt,shorten <=2.5pt] (0,-0.3) -- (.9,-.8);
      \draw[-,thin,shorten >=2.5pt,shorten <=2.5pt] (-.9,-1.4) -- (0,-1.9);
      \draw[-,thin,shorten >=2.5pt,shorten <=2.5pt] (.9,-1.4) -- (0,-1.9);
      \draw[-,thin,shorten >=2.5pt,shorten <=2.5pt] (0,-2.5) -- (0,-3);
    \end{tikzpicture} \qquad
    \label{fig:exdamonN}
    }~~\subfigure[$\Synt(P)$ in Example~\ref{ex:big}]{ \ \ \qquad
    \begin{tikzpicture}
      \node(1) at (0,0) [thick,rectangle,draw,inner sep=0cm,minimum height=.6cm,minimum width=.9cm] {$^*1$};
      \node(c) at (0,-1.1) [thick,rectangle,draw,inner sep=0cm,minimum height=.6cm,minimum width=.9cm] {$^*c$};
      \node(a) at (-.45,-2.2) [thick,rectangle,draw,inner sep=0cm,minimum height=.6cm,minimum width=.9cm] {$^*a$};
      \node(ac) at (.45,-2.2) [thick,rectangle,draw,inner sep=0cm,minimum height=.6cm,minimum width=.9cm] {$^*ac$};
      \node(b) at (-.45,-3.3) [thick,rectangle,draw,inner sep=0cm,minimum height=.6cm,minimum width=.9cm] {$^*b$};
      \node(bc) at (.45,-3.3) [thick,rectangle,draw,inner sep=0cm,minimum height=.6cm,minimum width=.9cm] {$^*bc$};
      \node(ab) at (-.45,-3.9) [thick,rectangle,draw,inner sep=0cm,minimum height=.6cm,minimum width=.9cm] {$^*ab$};
      \node(abc) at (.45,-3.9) [thick,rectangle,draw,inner sep=0cm,minimum height=.6cm,minimum width=.9cm] {$^*abc$};
      \draw[-,thin,shorten >=2.5pt,shorten <=2.5pt] (1) -- (0,-.8);
      \draw[-,thin,shorten >=2.5pt,shorten <=2.5pt] (0,-1.4) -- (0,-1.9);
      \draw[-,thin,shorten >=2.5pt,shorten <=2.5pt] (0,-2.5) -- (0,-3);
    \end{tikzpicture} \ \ \qquad
    \label{fig:exbig}
    }
  \caption{Egg-box representations}
  \label{fig:eggboxes}
  \end{figure}

  We have $P = [abc]$. The language $L= P^\omega = [abc]^\omega$ is weakly
  recognizable by $\Synt(P)$, too. All words in $\alpha \in L$ have
  infinitely many occurrences of the factor $ca$ and $\im(\alpha) =
  \Gamma$. In particular, $L$ is not open in the strict
  alphabetic topology. By Lemma~\ref{lem:DA:clopen}, the language
  $L$ is not strongly recognizable by any monoid in $\DA$.

  The conjugacy class of the linked pair $(abc,abc)$ is
  $\oneset{(ab,b),(ab,ab),(abc,bc),(abc,abc)}$ and $[abc,abc] = c^* a
  \Gamma^* b \Gamma^\infty \cap (\Gamma^* b)^\omega$. The language
  $[abc,abc]$ is strongly recognizable by $\Synt(P) \in \DA$. By
  Theorem~\ref{thm:fo2} it is $\FO^2$-definable. The set $[abc,abc]$
  is not open in the alphabetic topology. By Theorem~\ref{thm:delta2},
  $[abc,abc]$ is not $\Delta_2$-definable.

  The set $\overrightarrow{[abc]} = \overrightarrow{P} = c^* a
  \Gamma^* b \Gamma^\infty \cap (\Gamma^* c)^\omega$ is an arrow
  language which is weakly recognizable by $h$. It is not strongly
  recognized by the syntactic homomorphism of $P$ since
  $[abc][abc]^\omega \subseteq \overrightarrow{[abc]} \cap [abc,abc]$
  but $[abc,abc] \not\subseteq \overrightarrow{[abc]}$. On the other
  hand, $\overrightarrow{[abc]}$ is $\FO^2$-definable, and therefore,
  by Theorem~\ref{thm:fo2}, it is strongly recognizable by some other
  homomorphism onto a monoid in $\DA$.

  The $\gR$-class of $abc$ is ${\gR_{abc}} = \smallset{ab,abc}$. Hence
  $\overrightarrow{[\gR_{abc}]} = c^* a \Gamma^* b \Gamma^\infty$.  By
  Proposition~\ref{prop:arl} the complement of
  $\overrightarrow{[\gR_{abc}]}$ is also an arrow language; and by
  Theorem~\ref{thm:delta2} the language $\overrightarrow{[\gR_{abc}]}$
  is $\Delta_2$-definable.  Indeed, for $\overrightarrow{[\gR_{abc}]}$
  it is enough to say that there is some $b$ and there is some $a$
  with no $b$ to its left. This is a $\Sigma_2$-sentence. The
  equivalent $\Pi_2$-sentence says that there is some $b$ and for all
  $b$ there exists some $a$ to its left. It is also deterministic and
  \codet.
\end{example}

\section{Summary}\label{sec:summary}

We have given language-theoretic, algebraic and topological
characterizations for several first-order fragments over infinite
words. Since $\FO^2$ and $\Delta_2$ have the same expressive power
only when restricted to some fixed set of letters occurring infinitely
often (Thm.\,\ref{thm:lohengrin}), the picture becomes more complex
than in the case of finite words.  Among other results, we have shown
the relations in Figure~\ref{fig:bigpicture} between the fragments
$\FO^2$, $\Sigma_2$, $\Pi_2$, and $\Delta_2 = \Sigma_2 \cap \Pi_2$
(for completeness we included the fragments $\Sigma_1$, $\Pi_1$, their
Boolean closure $\mathbb{B}\Sigma_1$, and the Boolean closure
$\mathbb{B}\Sigma_2$ of $\Sigma_2$ in the picture).
\begin{figure}[ht]
\begin{center}
  \begin{tikzpicture}[scale=0.85]
    \draw[thin] (-.75,0) -- (12.75,0);
    \draw[thin,fill=DKbackcolora] (-.75,0) arc (180:0:6.75cm and 7.25cm);
    \draw[thin,fill=DKbackcolorA] (0,0) arc (180:0:4.75cm);
    \draw[thin,fill=DKbackcolorA] (12,0) arc (0:180:4.75cm);
    \draw[thin,fill=DKbackcolorA] (3.25,0) arc (180:0:2.75cm and 3.25cm);
    \draw[thin,fill=DKbackcolorA] (4.0,0) arc (180:0:1.25cm and 2.0cm);
    \draw[thin,fill=DKbackcolorA] (8.0,0) arc (0:180:1.25cm and 2.0cm);
    \draw[thin,fill=DKbackcolorA] (1.75,0) arc (180:0:4.25cm and 6.5cm);

    \begin{scope}
      \clip (0,0) arc (180:0:4.75cm);
      \draw[thin,fill=DKbackcolorB] (1.75,0) arc (180:0:4.25cm and 6.5cm);
    \end{scope}

    \begin{scope}
      \clip (12,0) arc (0:180:4.75cm);
      \draw[thin,fill=DKbackcolorB] (1.75,0) arc (180:0:4.25cm and 6.5cm);
    \end{scope}
    
    \begin{scope}
      \clip (12,0) arc (0:180:4.75cm);
      \draw[thin,fill=DKbackcolorC] (0,0) arc (180:0:4.75cm);
    \end{scope}

    \draw[thin,fill=DKbackcolorD] (3.25,0) arc (180:0:2.75cm and 3.25cm);
    \draw[thin,fill=DKbackcolorE] (4.0,0) arc (180:0:1.25cm and 2.0cm);
    \draw[thin,fill=DKbackcolorE] (8.0,0) arc (0:180:1.25cm and 2.0cm);

    \begin{scope}
      \clip (4.0,0) arc (180:0:1.25cm and 2.0cm);
      \draw[thin,fill=DKbackcolorF] (8.0,0) arc (0:180:1.25cm and 2.0cm);
    \end{scope}
    
    \draw[thick] (-.75,0) -- (12.75,0);
    \draw[thick] (-.75,0) arc (180:0:6.75cm and 7.25cm);
    \draw[thick] (0,0) arc (180:0:4.75cm);
    \draw[thick] (12,0) arc (0:180:4.75cm);
    \draw[thick] (3.25,0) arc (180:0:2.75cm and 3.25cm);
    \draw[thick] (4.0,0) arc (180:0:1.25cm and 2.0cm);
    \draw[thick] (8.0,0) arc (0:180:1.25cm and 2.0cm);
    \draw[thick] (1.75,0) arc (180:0:4.25cm and 6.5cm);

    \draw (6,7.14) node[below] {$\mathbf{\mathbb{B}\Sigma_2}$};
    \draw (0.75,0.25) node[above] {$\mathbf{\Sigma_2}$};
    \draw (11.25,0.25) node[above] {$\mathbf{\Pi_2}$};
    \draw (6,6.3) node[below] {\textbf{FO}$\mathbf{^2}$};
    \draw (6,4.4) node[below] {$\mathbf{\Delta_2}$};
    \draw (6,3.05) node[below] {$\mathbf{\mathbb{B}\Sigma_1}$};
    \draw (5.25,1.8) node[below] {$\mathbf{\Sigma_1}$};
    \draw (6.75,1.8) node[below] {$\mathbf{\Pi_1}$};

    \draw (1.25,2.0) node (L1) {$\mydot$};
    \draw (L1) node[above] {$\color{DKfontcolor}L_1$};

    \draw (2.5,2.0) node (L2) {$\mydot$};
    \draw (L2) node[above] {$\color{DKfontcolor}L_2$};

    \draw (4.9,6.5) node (L3) {$\mydot$};
    \draw (L3) node[above] {$\color{DKfontcolor}L_3$};

    \draw (6,5.0) node (L4) {$\mydot$};
    \draw (L4) node[above] {$\color{DKfontcolor}L_4$};

    \draw (4.9,3.3) node (L5) {$\mydot$};
    \draw (L5) node[above] {$\color{DKfontcolor}L_5$};

    \draw (6,1.95) node (L6) {$\mydot$};
    \draw (L6) node[above] {$\color{DKfontcolor}L_6$};

    \draw (4.9,0.5) node (L7) {$\mydot$};
    \draw (L7) node[above] {$\color{DKfontcolor}L_7$};

    \draw (6.9,0.5) node (L8) {$\mydot$};
    \draw (L8) node[above] {$\color{DKfontcolor}L_8$};

    \draw (9.5,2.0) node (L9) {$\mydot$};
    \draw (L9) node[above] {$\color{DKfontcolor}L_9$};

    \draw (10.75,2.0) node (L10) {$\mydot$};
    \draw (L10) node[above] {$\color{DKfontcolor}L_{10}$};

    \draw (6.03,0.50) node[above] (d1a) {$\color{DKfontcolor}\Gamma^{\infty}$};
    \draw (6,0.00) node[above] (d1b) {$\color{DKfontcolor}\emptyset$};
  \end{tikzpicture}
\parbox{3cm}{%
\begin{tabbing}
  aa \= $L_{11}$ \= \kill
  Here $\Gamma = \oneset{a,b,c}$ and \> \> \\[1mm]
  \> $L_1$ \> $= \text{``there exists a factor $ab$''} 
  \; = \; \Gamma^* ab \Gamma^{\infty}$, \\
  \> $L_2$ \> $= \text{``finitely many $a$'s''}
  \; = \; \Gamma^* \smallset{b,c}^\infty$, \\
  \> $L_3$ \> $= \text{``there is a factor $ab$ but no factor $ba$''}
  \; = \; L_1 \cap L_{10}$ \\
  \> $L_4$ \> $= \text{``finitely many $a$'s and infinitely 
    many $b$'s''} \; = \; L_2 \cap L_9$, \\
  \> $L_5$ \> $= \text{``the first $a$ occurs before the first $b$''}
  \; = \; c^* a \Gamma^* b \Gamma^{\infty}$, \\
  \> $L_6$ \> $= \text{``some $a$ occurs before some $b$ but 
    no $a$ occurs before any $c$''}$ \\
  \> \> $= \,\Gamma^* a \Gamma^* b \Gamma^{\infty} \;\cap\; 
  \smallset{b,c}^* a \smallset{a,b}^\infty \; = \; L_7 \cap L_8$, \\
  \> $L_7$ \> $= \text{``some $a$ occurs before some $b$''} 
  \; = \; \Gamma^* a \Gamma^* b \Gamma^{\infty}$, \\
  \> $L_8$ \> $= \text{``no $a$ occurs before any $c$''}
  \; = \; \smallset{b,c}^* a \smallset{a,b}^\infty 
  \; \cup \; \smallset{b,c}^\infty$, \\
  \> $L_9$ \> $= \text{``infinitely many $b$'s''} 
  \; = \; (\Gamma^* b)^\omega$, \\
  \> $L_{10}$ \> $= \text{``there is no factor $ba$''}
  \; = \; \Gamma^{\infty} \setminus \Gamma^* ba \Gamma^{\infty}$.
\end{tabbing}}
\vspace{-1.5\baselineskip}
\end{center}
\caption{Some examples and the relations between the different
  fragments}
\label{fig:bigpicture}
\end{figure}
The intersection $\Delta_1 = \Sigma_1 \cap \Pi_1$ contains only the
trivial languages $\emptyset$ and $\Gamma^{\infty}$.  The language
$L_9$ in Figure~\ref{fig:bigpicture} is the closure of $L_4$ within
the alphabetic topology.  The interior of $L_4$ (as well as of any
other language in $\Gamma^\oo$) with respect to the alphabetic
topology is empty.  Another example in $\Sigma_2 \cap \FO^2$ which is
not in $\Pi_2$ is the set of all finite words
$\Gamma^*$. Symmetrically, the language of all infinite words
$\Gamma^\omega$ is in $\Pi_2 \cap \FO^2$ but not in $\Sigma_2$.

In order to sketch the main results on small first-order fragments
over finite and infinite words in Table~\ref{tab:main}, we introduce
some terminology. By $\mathrm{Pol}$ we denote the language class of
polynomials, $\mathrm{UPol}$ are unambiguous polynomials, and
\emph{restricted} $\mathrm{UPol}$ is a proper subclass of
$\mathrm{UPol}$. \emph{Simple polynomials} are finite unions of
languages of the form $\Gamma^* a_1 \cdots \Gamma^* a_n
\Gamma^\infty$. A language $L \subseteq \Gamma^\infty$ is
\emph{piecewise testable} if there exists some $k \in \N$ such that
for every $\alpha \in \Gamma^\infty$ membership in $L$ only depends on
the set of scattered subwords of $\alpha$ of length less than $k$. The
first-order fragment $\Sigma_1$ consists of first-order sentences in
prenex normal without universal quantifiers; its negations are in
$\Pi_1$ and its Boolean closure is $\mathbb{B}\Sigma_1$.

All of the algebraic properties in Table~\ref{tab:main} are decidable,
since the syntactic monoid of a regular language is effectively
computable~\cite{pp04,tho90handbook}. Together with the
$\mathrm{PSPACE}$-completeness of the problem whether a language is
closed in the alphabetic topology (Thm.\,\ref{thm:pspaceclosure}),
this yields decidability of the membership problem for the respective
first-order fragments as a corollary. Decidability was shown before by
Wilke~\cite{wil98} for $\FO^2$ and by Boja{\'n}czyk
\cite{boj08fossacs} for $\Sigma_2$. The characterization for the
fragment $\Sigma_1$ is due to Pin~\cite{Pin98a}; see also~\cite{pp04}.
The same holds for the Boolean closure of $\Sigma_1$ except for the
topological part of the ``Algebra $+$ Topology'' characterization,
which is a consequence of Corollary~\ref{cor:clopen:det}.

\begin{table}[ht]
\begin{center}
  \begin{tabular}{ccrcll}
    \toprule
    \textbf{Logic} & \textbf{Languages} & \textbf{Algebra} 
    & \textbf{+}  & \textbf{Topology} & \\
    \toprule
    $\Sigma_2$ &
    $\mathrm{Pol}$ &
    $e M_e e \leq e$ & 
    $+$ &
    \parbox{3cm}{
      open \\ alphabetic} &
    Thm.\,\ref{thm:sigma2} \\
    \cmidrule{1-6}
    $\Pi_2$ & &
    $e M_e e \geq e$ &
    $+$ &
    \parbox{3cm}{
      closed \\ alphabetic} &
    Thm.\,\ref{thm:sigma2} \\
    \cmidrule{1-6}
    \multirow{2}{15mm}{\centering$\FO^2$} &
    \multirow{2}{25mm}{\centering$\mathrm{UPol}$ $+$ $A^{\im}$} &
    $\DA$ & & &
    \multirow{2}{15mm}{Thm.\,\ref{thm:fo2}} \\
     & & weak $\DA$ &
    $+$ &
    \parbox{3.5cm}{
      closed \\ strict alphabetic} & \\
    \cmidrule{1-6}
    $\FO^2 \cap \Sigma_2$ &
    $\mathrm{UPol}$ &
    $\DA$ &
    $+$ &
    \parbox{3cm}{
      open \\ alphabetic} &
    Thm.\,\ref{thm:fo2si2} \\
    \cmidrule{1-6}
    $\FO^2 \cap \Pi_2$ &
    &
    $\DA$ &
    $+$ &
    \parbox{3cm}{
      closed \\ alphabetic} &
    Thm.\,\ref{thm:fo2pi2} \\
    \cmidrule{1-6}
    $\Delta_2$ &
    restricted $\mathrm{UPol}$ &
    $\DA$ &
    $+$ &
    \parbox{3cm}{
      clopen \\ alphabetic} &
    Thm.\,\ref{thm:delta2} \\
    \cmidrule{1-6}
    $\mathbb{B}\Sigma_1$ &
    piecewise testable &
    $\gJ$-trivial &
    $+$ &
    \parbox{3cm}{
      clopen \\ alphabetic} &
    \parbox{1.5cm}{
      Cor.\,\ref{cor:clopen:det} \\
      and~\cite{pp04}} \\
    \cmidrule{1-6}
    $\Sigma_1$ &
    simple $\mathrm{Pol}$ &
    $x \leq 1$ &
    $+$ &
    \parbox{3cm}{
      open \\ Cantor} &
    \cite{pp04} \\
    \cmidrule{1-6}
    $\Pi_1$ & &
    $x \geq 1$ &
    $+$ &
    \parbox{3cm}{
      closed \\ Cantor} &
    \cite{pp04} \\
    \bottomrule
  \end{tabular}
\end{center}
\caption{Main characterizations of some first-order fragments}
\label{tab:main}
\end{table}

\section{Outlook and open problems}

By definition, $\Sigma_1$-definable languages are open in the Cantor
topology. We introduced an alphabetic topology such that
$\Sigma_2$-definable languages are open in this topology. Therefore,
an interesting question is whether it is possible to extend this
topological approach to higher levels of the first-order alternation
hierarchy.  To date, even over finite words no decidable
characterization of the Boolean closure of $\Sigma_2$ is known.  In
case that a decidable criterion is found, it might lead to a decidable
criterion for infinite words simply by adding a condition of the form
``$L$ and its complement are in the second level of the Borel
hierarchy of the alphabetic topology''. Another possible way to
generalize our approach might be combinations of algebraic and
topological characterizations for fragments with successor predicate
$\mathrm{suc}$ such as $\FO^2[{<},\mathrm{suc}]$ or
$\Sigma_2[{<},\mathrm{suc}]$. A characterization of those languages
which are weakly recognizable by monoids in $\DA$ is also open.

\section*{Acknowledgements}

We thank the anonymous referees for many useful suggestions which helped to improve 
the presentation of the paper. We also 
thank Luc Dartois from the ENS Cachan for helpful discussions during
his internship in Stuttgart in 2008 and for his firm conviction that
$\FO^2$ should coincide with $\DA$.

%%%%%%%%%%%%%%%%%%%%%%%%%%%%%%%%%%%%%%%%%%%%%%%%%%%%%

%\bibliographystyle{abbrv}
%\bibliography{traces}

\newcommand{\Ju}{Ju}\newcommand{\Ph}{Ph}\newcommand{\Th}{Th}\newcommand{\Yu}{Y%
u}

%%%%%%%%%%%%%%%%%%%%%%%%%%%%%%%%%%%%%%%%%%%%%%%%%%%%%

\end{document}